\documentclass[pra,final,twocolumn,showpacs,showkeys]{revtex4}
\usepackage{amsmath}
\usepackage{graphicx}
\usepackage{amsfonts}
\usepackage{amssymb}
\usepackage{subfigure}

\newcommand{\subs}[1]
{ 
	\mbox{\scriptsize{#1}}
}

\begin{document}



\title{Quantum logic gates for superconducting resonator qudits}

\author{Frederick W. Strauch} \email[Electronic address: ]{Frederick.W.Strauch@williams.edu}
\affiliation{Williams College, Williamstown, MA 01267, USA}
\date{\today}

\begin{abstract}
We study quantum information processing using superpositions of Fock states in superconducting resonators, as quantum $d$-level systems (qudits).  A universal set of single and coupled logic gates is theoretically proposed for resonators coupled by superconducting circuits of Josephson juctions.  These gates use experimentally demonstrated interactions, and provide an attractive route to quantum information processing using harmonic oscillator modes.

\end{abstract}
\pacs{03.67.Bg, 03.67.Lx, 85.25.Cp}
\keywords{Qubit, entanglement, quantum computing, superconductivity, Josephson junction.}
\maketitle

\section{Introduction}

Superconducting quantum bits (qubits) \cite{Clarke2008} are a leading candidate for a solid-state quantum computer.  However, while coherence times are continually increasing, it remains necessary to study how to maximize coherence while accessing the large Hilbert space required by key applications in quantum information processing.   Examples include the rapid controlled-phase gate using auxiliary states \cite{DiCarlo2009,DiCarlo2010,Yamamoto10,Strauch2003} and the general framework to improve quantum logic gate synthesis using multilevel systems \cite{Lanyon09}.  An emerging pattern is that resources outside of the traditional qubit states can lead to improved control sequences with reductions in total time or complexity.

Superconducting phase and transmon circuits are a natural candidate to explore operations outside of the qubit subspace, as these systems are in fact weakly anharmonic oscillators with many levels.  Control of multiple levels in these devices has been demonstrated experimentally \cite{Claudon2004,Dutta2008,Neeley2009,Bianchetti10} and explored theoretically \cite{Tian2000,Steffen2003,Amin2006,Strauch2007,Forney10}.   Notably, a theoretical method \cite{Motzoi09} incorporating multiple levels has led to improvements in qubit logic operations \cite{,Lucero10,Chow10}.  

Superconducting resonators can also be controlled at the Fock state level.   By coupling such resonators to an auxiliary nonlinear system, recent experiments have created Fock states \cite{Hofheinz2008}, observed their decay \cite{Wang2008}, and demonstrated the synthesis of arbitrary superpositions of Fock states \cite{Hofheinz2009}.  This last experiment used the protocol of Law and Eberly \cite{Law96} with linear coupling of a phase qubit to a coplanar waveguide resonator.  A recent theoretical work \cite{Strauch10} extended this approach to the synthesis of entangled states of two (and possibly more) resonators.  Subsequently, an experimental synthesis \cite{wang11} of a ``high'' NOON state \cite{Dowling08} was accomplished using an alternative procedure \cite{Merkel10}.

While great progress has thus been made in the control of superconducting resonators, these works leave open the question of whether the larger state space of the resonator can be used to process quantum information.  While there are certainly caveats to the question of ``qubit or oscillator?'' (see, e.g. \cite{mikeandike}), there is an established body of work demonstrating that quantum systems with multiple states, known as qudits (for $d$-level systems), can be as useful as qubits.  Using the lowest $d$ levels of a harmonic oscillator would thus be a potential alternative to qubits.  For superconducting circuits in particular, it is clear that resonators can be fabricated with much greater precision and coherence, and thus a central question is how to compute using the additional resources present in harmonic oscillator modes.

An important step in that direction was taken by Jacobs \cite{Jacobs07x}, who showed that linear coupling of an oscillator to an auxiliary qubit was sufficient to approximate any desired evolution of the oscillator.  This was based on the general Lie algebraic result by Lloyd {\it{et al.}} \cite{Lloyd04}, but applying this to quantum logic on a discrete set of Fock states would require significant overhead in complexity (to synthesize the desired interactions).  Inducing a nonlinearity perturbatively \cite{Jacobs09c} is another route to unitary control of the oscillator, although this may require some compromise in timescales (to stay within the perturbative limits).   A scheme of this sort, appropriate to atomic cavity-QED or ion trap systems, was proposed by Santos \cite{Santos05} and serves as a primary inspiration for our proposal. Here we present a detailed analysis of a circuit-QED approach in which a three-level system, such as a phase or transmon qubit, is used as an auxiliary to enable arbitrary unitary control of a superconducting resonator.

\begin{figure}[t]
\begin{center}
\includegraphics[width=3in]{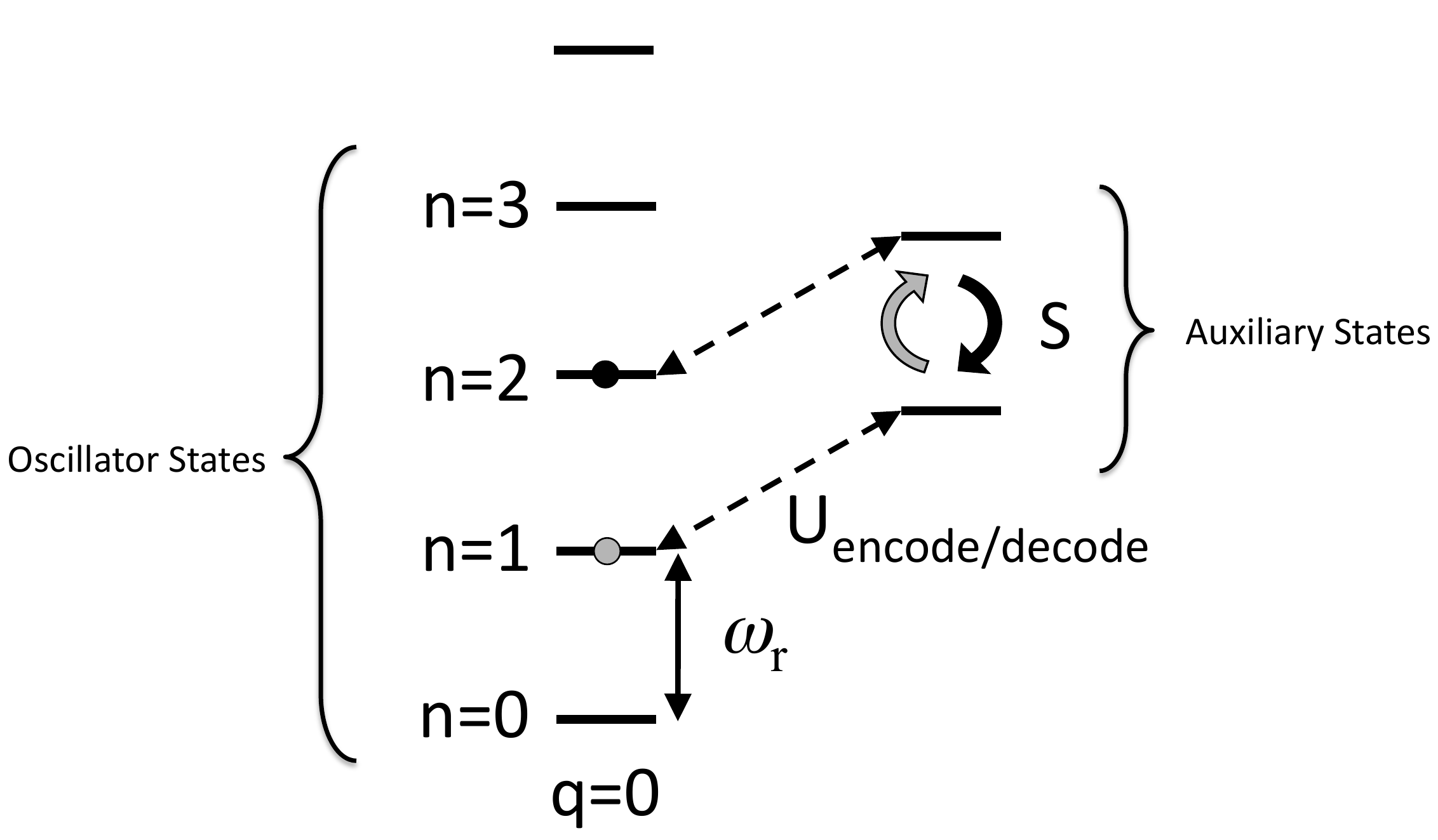}
\end{center}
\caption{General approach to quantum logic using harmonic oscillator states.  Encoding and decoding operations transfer two oscillator states (here $n=1$ and $n=2$) to a pair of auxiliary states which can be swapped by $S$.  Here the total single-qudit rotation is $U_{1,2} = U_{\subs{decode}} S U_{\subs{encode}}$}. 
\label{fig1}
\end{figure}

In this work we combine two experimentally demonstrated interactions to propose a simple procedure to perform an arbitrary rotation between Fock states, and by composition an arbitrary unitary operation on the Fock states.  The basic idea is shown in Fig. \ref{fig1}.  We devise a control sequence to selectively move two states of the oscillator to auxiliary levels.  Here a rotation or swap $S$ is performed between these two auxiliary levels, and finally these levels are returned to the original oscillator states.  For convenience, we will call the first step an encoding operation $U_{\subs{encode}}$ and the final step a decoding operation $U_{\subs{decode}}$, so that the net rotation is  $U_{1,2} = U_{\subs{decode}} S U_{\subs{encode}}$.  

The first ingredient in our proposal is a quasi-dispersive interaction between a qubit and the resonator, to allow for number-state-dependent rotations of the qubit.  This was first seen spectroscopically by Schuster {\it et al.} \cite{Schuster07}, and more recently used to perform a non-demolition measurement of a resonator memory by Johnson {\it et al.} \cite{Johnson10}.  We propose to use this interaction to selectively address the Fock states of interest as part of the encoding and decoding operations.  This approach was previously used in the entangled state synthesis algorithm \cite{Strauch10}.

The second ingredient is a resonant swapping interaction between the resonator and the higher levels of a superconducting phase or transmon qubit.  This was used in the NOON-state synthesis experiment \cite{wang11}, and effects the qudit rotation by swapping the auxiliary levels, as shown in Fig. 1.

This paper is organized as follows.  In Section II we briefly review existing qudit theory and outline how our scheme can be used for qudit logic operations.  In Section III, the basic system of a three-level system coupled to an oscillator is presented and analyzed.  In Section IV the time-dependent control sequences are described and verified by numerical simulations.  In Section V we show how this can be extended to a two-qudit logic gate.  In Section VI we analyze the effects of decoherence and discuss resonator measurement.  Finally, we conclude in Section VII with a discussion of open topics for study.

\section{Qudit Logic}

Multilevel quantum logic has been explored as an alternative to the traditional qubit constructions by many authors \cite{Gottesman99, Gottesman01, Stroud2000, Bartlett02}.  We follow the discussion by Brennen {\it et al.} \cite{Brennen05}.  They show, using the QR decomposition from linear algebra, that arbitrary single-qudit unitaries can be constructed from a family of two-component rotations
\begin{equation}
U_{j,k}(\lambda, \phi) = \mathcal{R}_{jk}^{z} (\phi) \mathcal{R}_{jk}^{x} (2\lambda) \mathcal{R}_{jk}^{z} (-\phi)
\end{equation}
where we have defined two operations in the qudit subspace $\{|j\rangle, |k\rangle\}$:
\begin{equation}
\mathcal{R}_{j,k}^{x} (\theta) = \exp \left[- i \frac{\theta}{2} \left( |j\rangle \langle k| + |k \rangle \langle j| \right) \right]
\label{singlex1}
\end{equation}
and
\begin{equation}
\mathcal{R}_{j,k}^{z} (\theta) = \exp \left[ - i \frac{\theta}{2} \left( |j\rangle \langle j| - |k\rangle \langle k| \right) \right].
\label{singlex2}
\end{equation}
In addition to these single-qudit rotations, we also need a two-qudit operation, to generalize the controlled-NOT gate commonly used in qubit circuits.  We shall synthesize the controlled-phase gate
\begin{equation}
\mathcal{U}_{j,k} (\theta) = \exp \left(- i \theta |j,k\rangle \langle j,k| \right),
\label{cphasex}
\end{equation}
where $|j,k\rangle = |j\rangle \otimes |k\rangle$ is the state in which the first qudit is in state $|j\rangle$ and the second qudit is in state $|k\rangle$.  
This gate set is sufficient to perform an arbitrary two-qudit unitary operation \cite{Brennen05}, and by extension to multiple qudits, universal quantum computation \cite{Stroud2000}. Explicit constructions for circuit synthesis can be found in \cite{Bullock05,OLeary06}.  

In the implementation we will present shortly, the rotations will be between neighboring oscillator states $j$ and $k=j+1$.   That is, we will construct $U_{j,j+1}$ from the sequence $U_{\subs{decode}} S U_{\subs{encode}}$, as illustrated in Fig. 1, where $S$ performs a swapping interaction between the amplitudes for states $|j\rangle$ and $|j+1\rangle$.  Note that this limitation to neighboring oscillator states does not present a true obstacle to general qudit logic.  As shown in Lemma II.1 of \cite{Brennen05}, the important requirement is that there is a connected coupling graph between the qudit states.  Rotations between neighboring states leads to a linear coupling graph: 
\begin{equation}
0 \leftrightarrow 1 \leftrightarrow 2 \leftrightarrow \cdots \leftrightarrow d-1.
\end{equation} 

Finally, the single-qudit phase rotations $R_{j,k}^{z}$ can be performed as in current experiments \cite{DiCarlo2009,DiCarlo2010}, by short detuning pulses that can be incorporated in the single-qudit rotations.    

\section{Superconducting Implementation}

\begin{figure}[t]
\begin{center}
\includegraphics[width=2.5in]{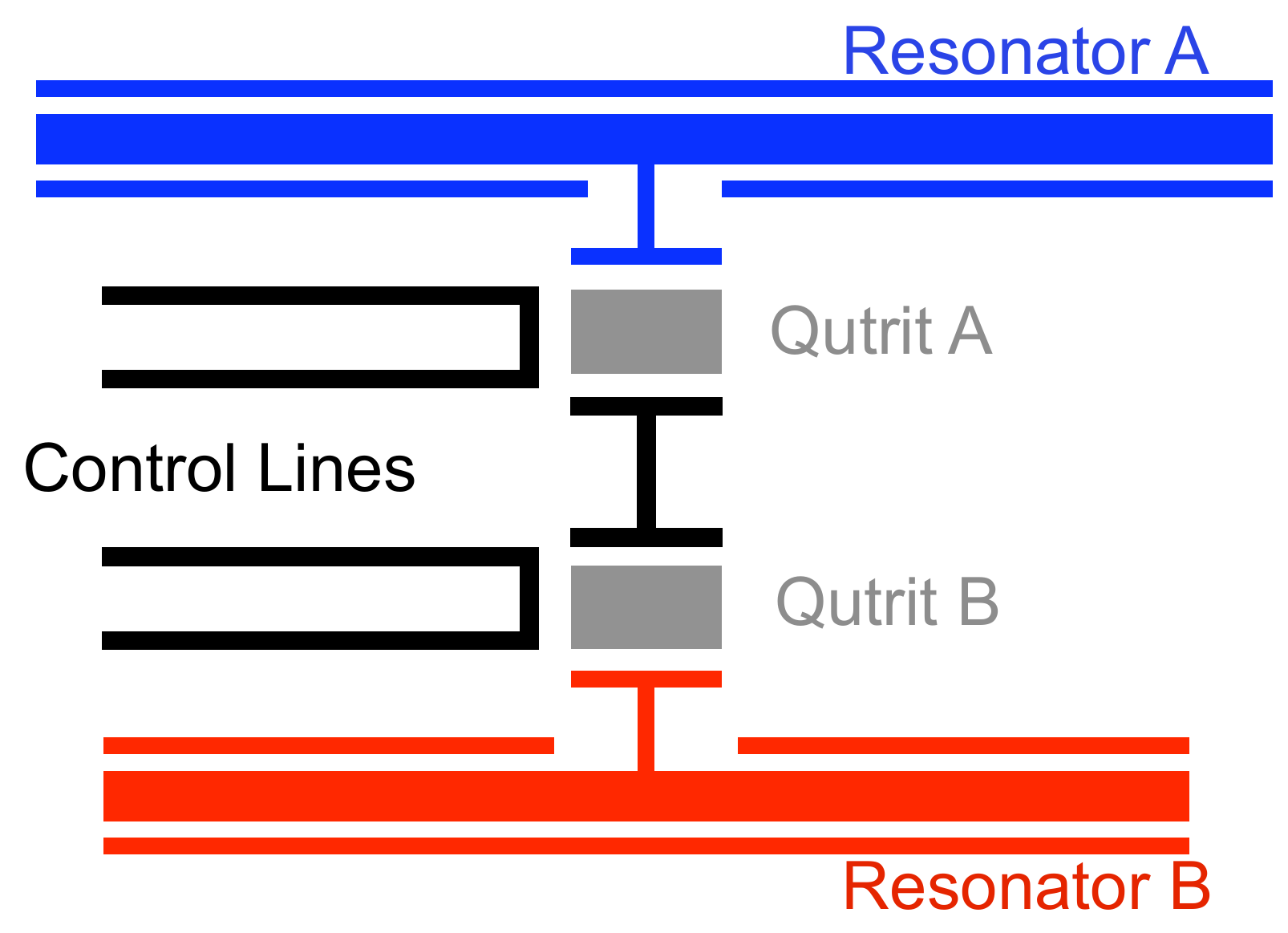}
\end{center}
\caption{Schematic superconducting circuit to implement single and two-qudit operations on resonators $A$ and $B$.  Each resonator is coupled to an auxiliary qutrit ($A$ or $B$), which are themselves coupled to each other.  Control lines allow manipulation of the qutrits.}
\label{fig2}
\end{figure}

We extend the framework of \cite{Strauch10}, in which two superconducting resonators are coupled by a tunable circuit, as shown in Fig. \ref{fig2}.  
Letting $a$ and $a^{\dagger}$ be the ladder operators for resonator $A$, $b$ and $b^{\dagger}$ for resonator $B$, we model the system as 
\begin{eqnarray}
\mathcal{H} &=& \mathcal{H}_{\subs{A}}  + \hbar \omega_a a^{\dagger} a + \hbar g_a \left(a \sigma_+^{A} + a^{\dagger} \sigma_-^{A} \right) \nonumber \\
& & + \ \mathcal{H}_{\subs{B}} + \hbar \omega_b b^{\dagger} b + \hbar g_b \left(b \sigma_+^{B} + b^{\dagger} \sigma_-^{B} \right) \nonumber \\
& & + \ \hbar g_{ab} \left(\sigma_{+}^{A} \sigma_{-}^{B} + \sigma_{-}^{A} \sigma_{+}^{B} \right).
\end{eqnarray}
Here $\mathcal{H}_{\subs{A}}$ and $\mathcal{H}_{\subs{B}}$ are the single-qubit Hamiltonians for the auxiliary, and $\sigma_{\pm}^{A}$ and $\sigma_{\pm}^{B}$ are the corresponding raising and lowering operators (see below).  We will assume the coupling $g_{ab}$ between the two auxiliaries can be turned on and off at will, using the tunable coupling circuits recently demonstrated \cite{Allman2010,Bialczak10}, and that the auxiliaries can be controlled by microwave and flux pulses.

\subsection{Single Resonator Model}

We begin by focusing on a single qubit-resonator system (i.e. $A$ or $B$), described by the following Hamiltonian 
\begin{equation}
\mathcal{H} = \mathcal{H}_{\subs{Q}} + \hbar \omega_r a^{\dagger} a + \hbar g \left(a \sigma_+ + a^{\dagger} \sigma_- \right),
\end{equation}
where the auxiliary quantum system is taken as a three-level qutrit. 
\begin{equation}
\mathcal{H}_{\subs{Q}} = \hbar \left(\begin{array}{ccc} 
0 & 0 & 0 \\
0 & \omega_{01} & 0 \\
0 & 0 & \omega_{02} 
\end{array} \right)
\end{equation}
and
\begin{equation}
\sigma_- = \left(\begin{array}{ccc} 0 & 1 & 0 \\
0 & 0 & \lambda \\
0 & 0 & 0 \end{array} \right),
\end{equation}
with $\sigma_+ = \sigma_-^{\dagger}$.  Note that this auxiliary system could be either a phase or transmon qubit, as each have a similar level structure, in that $\omega_{12} = \omega_{02} - \omega_{01} < \omega_{01}$, and $\lambda \approx \sqrt{2}$.  In addition, they are both tunable by external flux pulses, which we will use in our construction.

\begin{figure}
\begin{center}
\includegraphics[width=3in]{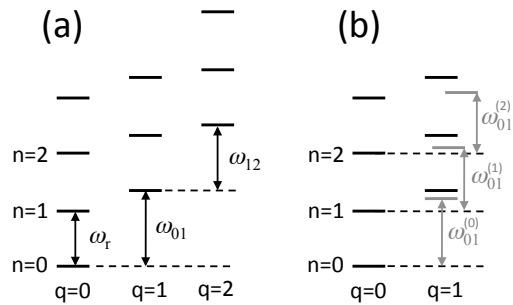}
\end{center}
\caption{ (a) Energy level diagram for a three-level artificial atom coupled to a resonator.  The artificial atom has level spacings $\omega_{01}$ and $\omega_{12} < \omega_{01}$, while the resonator has frequency $\omega_r$  (b) Approximate energy level diagram in the dispersive regime, in which the second excited state of the atom has been eliminated.  The dressed states have number-state-dependent level spacings $\omega_{01}^{(n)}$. }
\label{level0}
\end{figure}

An energy level diagram is shown in Fig. \ref{level0}(a), where we have used the convention to label the system by $|q,n\rangle$, were $q=0,1,2$ is the state of auxiliary qutrit and $n$ is the photon number (or Fock state).  This is a generalization of the classic Jaynes-Cummings Hamiltonian to a three-level artificial atom coupled to a resonator.  

This Hamiltonian above conserves the excitation number
\begin{equation}
N = a^{\dagger} a + \left( |1\rangle \langle 1| + 2 |2\rangle \langle 2| \right),
\end{equation}
so that we can break the problem into an infinite set of (up to) $3 \times 3$ blocks.  Using the notation $|q,n\rangle$ for qubit states $q=0,1,2$ and Fock states $n$, the ground state $|q=0,n=0\rangle$ is unique and set to have zero energy.  The first-excited subspace of $|q=0,n=1\rangle$ and $|q=1,n=0\rangle$ is governed by the Hamiltonian
\begin{equation}
\mathcal{H}_1 = \hbar \left( \begin{array}{cc}
\omega_{r} & g \\
g & \omega_{01}
\end{array} \right).
\end{equation}
The remaining states involve $3 \times 3$ matrices for the states $|0,n\rangle$, $|1,n-1\rangle$, $|2,n-2\rangle$: 
\begin{equation}
\mathcal{H}_2 = \hbar \left( \begin{array}{ccc}
n \omega_r & g \sqrt{n} & 0 \\
g \sqrt{n} & (n-1) \omega_r + \omega_{01} & g \lambda \sqrt{n-1} \\
0 & g \lambda \sqrt{n-1} & (n-2) \omega_r + \omega_{02} 
\end{array} \right).
\end{equation}

Assuming that we are away from the avoided crossings $\omega_{01} = \omega_r$, $\omega_{12} = \omega_r$ or $\omega_{02} = 2 \omega_r$, we can apply perturbation theory to $\mathcal{H}_2$ to find the following for the energies $E_{q,n}$:
\begin{eqnarray}
E_{0,n}/\hbar &\approx& n \omega_r + n \frac{g^2}{\omega_r - \omega_{01}}, \\
E_{1,n}/\hbar &\approx& n \omega_r + \omega_{01} + (n+1)  \frac{g^2}{\omega_{01} - \omega_r} \nonumber \\
& & \quad \quad + n \frac{g^2 \lambda^2}{\omega_r - \omega_{12}}, \\
E_{2,n}/\hbar &\approx& n \omega_r + \omega_{02} + (n+1) \frac{g^2 \lambda^2}{\omega_{12} - \omega_{r}}.
\end{eqnarray}
The shift of the eigenvalues is the AC Stark shift, and have been seen for coupling of a qubit to both quantum and classical fields.  In the dispersive regime, we can effectively eliminate state $q=2$ to have the modified level diagram shown in Fig.~\ref{level0}(b). 

As a consequence of this shift, the transition between qubit states depends on the photon number. Defining
\begin{equation}
\omega_{01}^{(n)} = \frac{E_{1,n} - E_{0,n}}{\hbar} = \omega_{01} + \frac{g^2}{\omega_{01} - \omega_r} (2n+1) + \frac{g^2 \lambda^2}{\omega_r - \omega_{12}} n. 
\end{equation}
This shift of the qubit transition is indicated in Fig.~\ref{level0}(b).  This Stark shift can be used to provided a number-state-dependent transition, effectively a controlled-rotation of the qubit based on the Fock state of the resonator, by applying an additional microwave field to the qubit of the form $\mathcal{H_{\subs{drive}}}~=~\hbar\Omega (\sigma_+ + \sigma_-) \cos \omega t$. 

\begin{figure}
\begin{center}
\includegraphics[width=3in]{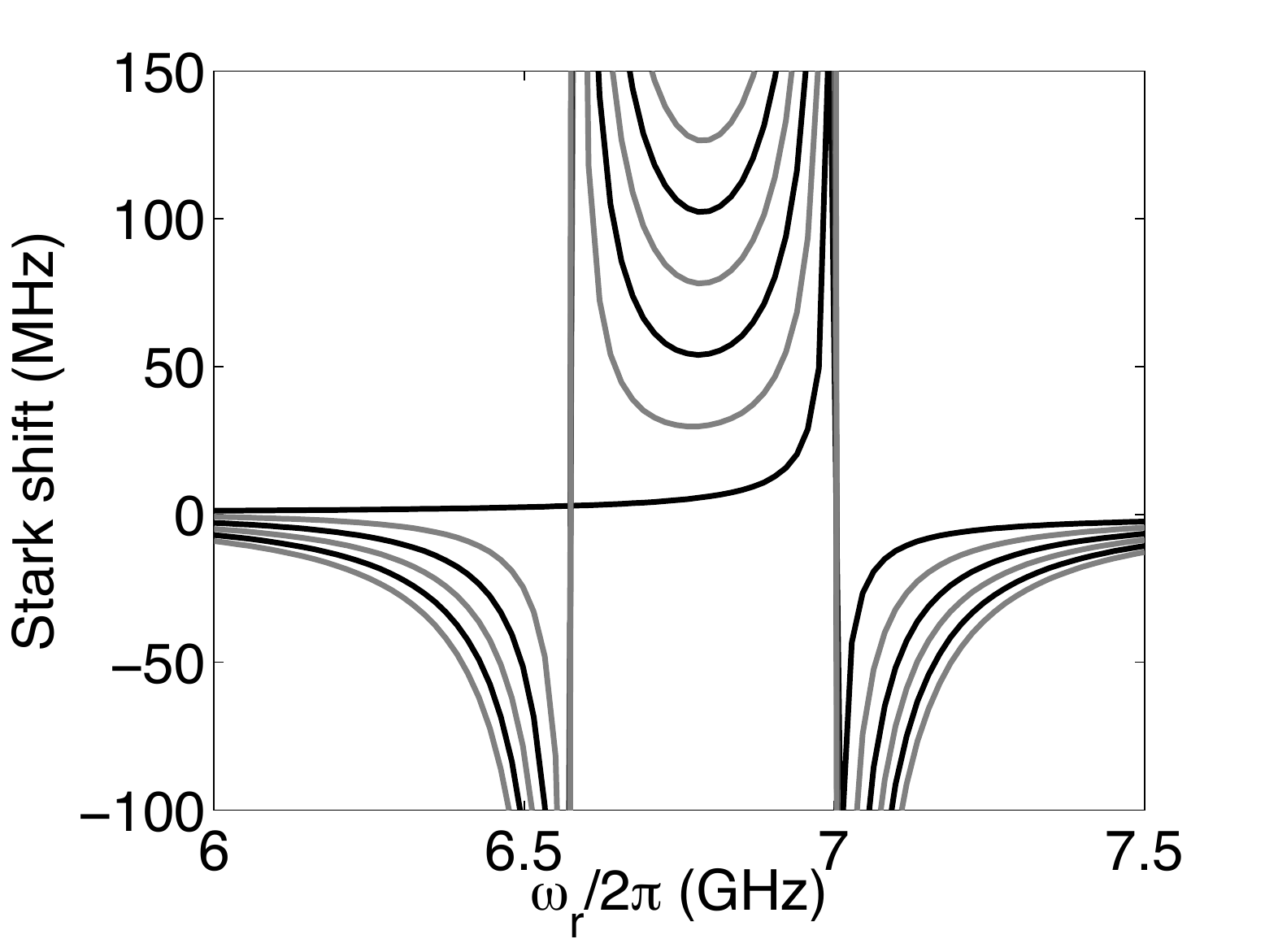}
\end{center}
\caption{ The number-state-dependent Stark shift $\omega_{01,n}-\omega_{01}$ as a function of the resonator frequency $\omega_r$.  The perturbative Stark shift is shown for $n=0 \to 5$ as a function of the resonator frequency $\omega_r/2\pi$, for typical qubit parameters $\omega_{01}/2\pi = 7 \ \mbox{GHz}$, $\omega_{12}/2\pi = 6.58 \ \mbox{GHz}$, $\lambda = 1.46$ and coupling $g/2\pi= 35 \ \mbox{MHz}$.}
\label{stark1}
\end{figure}

The frequency shift  $\omega_{01}^{(n)} - \omega_{01}$ is shown in Fig.~\ref{stark1}, with typical experimental parameters.  As described in Koch {\it et al.} \cite{Koch07}, there are three special regions in this figure: $\omega_r < \omega_{12}$, $\omega_{12} < \omega_r < \omega_{01}$, and $\omega_{01} < \omega_r$.  This is different from what would be expected for a resonator coupled to a two-level system, which would have only two regions, one with positive shift and one with negative shift.  The middle region with positive shift is known as the ``straddling'' regime, and has the largest value, while the negative regions have smaller shifts.  
The divergences in Fig.~\ref{stark1} are at the resonant conditions $\omega_r = \omega_{01}$ or $\omega_r = \omega_{12}$.  These are in fact avoided crossings where the states (and transitions between them) are more complicated. 

These avoided crossings can also be used for control.  By rapidly shifting the qubit frequency to one of these anticrossings by a ``shift'' pulse, swapping between the hybridized states occurs \cite{Strauch2003}.  This has been used to swap excitations from the qubit to the oscillator \cite{Sillanpaa2007} and to prepare Fock states and their superpositions \cite{Hofheinz2008,Hofheinz2009}.  We will consider the anticrossing at $\omega_{12} = \omega_r$ (see the next section), as was recently used for NOON state prepration \cite{wang11}.

\subsection{Qudit operation}

\begin{figure}
\begin{center}
\includegraphics[width=3in]{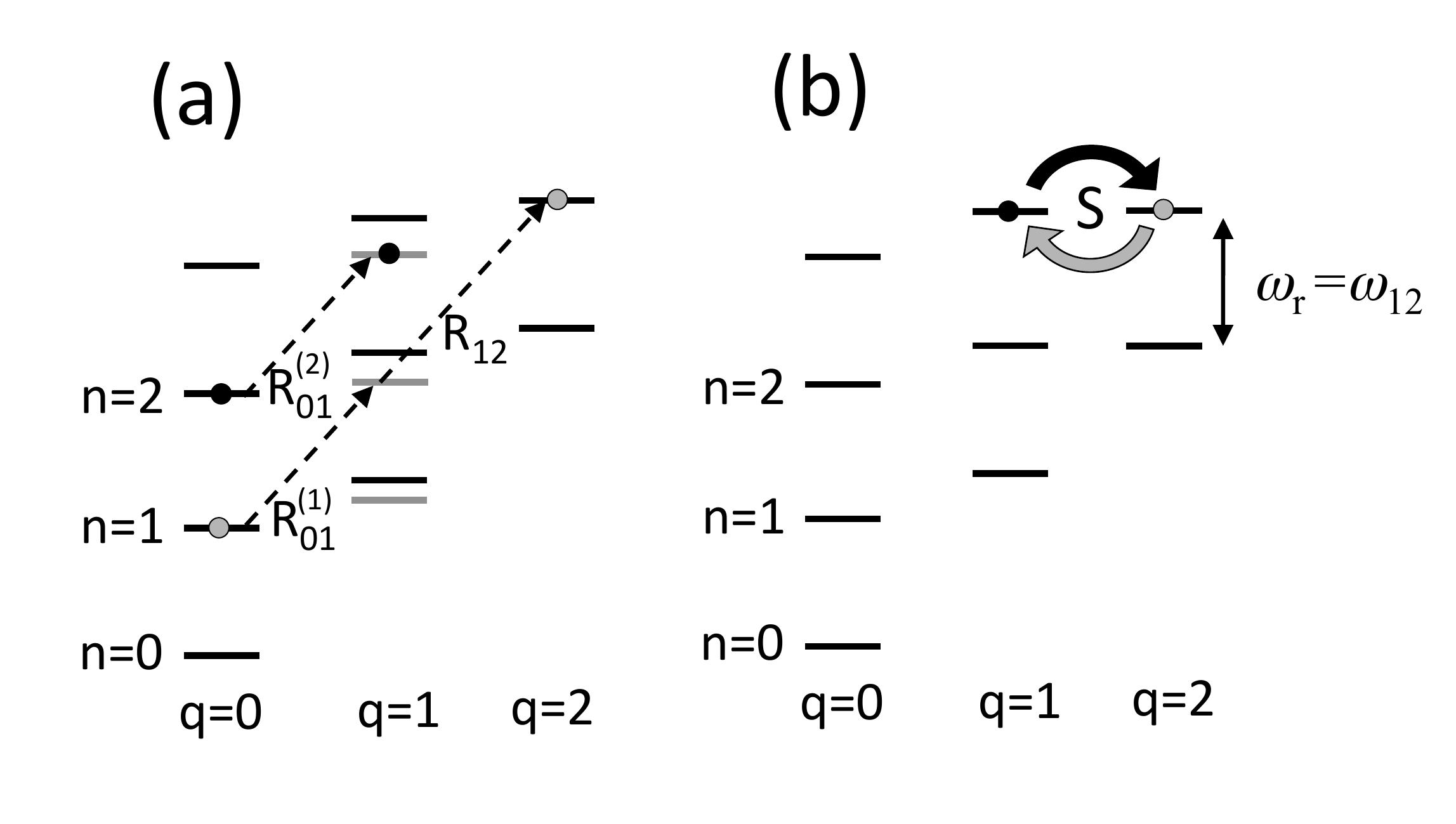}
\end{center}
\caption{(a) Encoding and decoding sequence for resonator logic gate.  The number-state-dependent qubit rotations $R_{01}^{(1)}$ and $R_{01}^{(2)}$, in the dispersive regime, select out the Fock states $n=1$ and $n=2$, while the $R_{12}$ transition prepares the state for the logic gate.  (b) The swap gate $S$ performs the logic gate in the resonant regime $\omega_{12} = \omega_r$. }
\label{gate}
\end{figure}

Having defined the quantum system, we now illustrate how the dispersive and resonant interactions can be used for arbitrary single-qudit operation.  Consider a quantum state:
\begin{equation}
|\psi_0\rangle = |0\rangle \otimes \sum_{n=0}^{d-1} c_n |n\rangle.
\end{equation} 
We begin by performing a rotation between neighboring Fock states $j$ and $j+1$.  To do this, we first apply a number-state-dependent $\pi$-pulse, conditioned on the photon state $n=j$, performing the transformation $|0,n\rangle \to |\delta_{n,j},n\rangle$; this will be called $R_{01}^{(j)}$.  This is followed by a $\pi$-pulse on the $q=1 \to 2$ transition, called $R_{12}$, after which another number-state selective $\pi$ pulse is performed, conditioned on the photon state $n=j+1$.  The net result of these operations is to transform $|\psi_0\rangle$ into
\begin{eqnarray}
|\psi_1\rangle &=& U_{\subs{encode}} |\psi_0\rangle \nonumber \\
&=& c_j |2,j\rangle + c_{j+1} |1,j+1\rangle +  \sum_{n \ne j,j+1} c_n |0,n\rangle, \nonumber \\
\end{eqnarray}
where $U_{\subs{encode}} = R_{01}^{(j+1)} R_{12} R_{01}^{(j)}$.  This has selected out the $|j\rangle, |j+1\rangle$ subspace of the resonator; this sequence is illustrated in Fig. \ref{gate}(a) for $j=1$.  

The system is now configured to the resonant regime, with the resonator frequency $\omega_r$ equal to the $q=1\to2$ transition frequency $\omega_{12}$, as shown in Fig.~\ref{gate}(b).  This can be done by dynamically tuning the qubit frequency, and was the key step in the NOON state experiment \cite{wang11}.  The subsequent evolution is a two-state oscillation between $|1,j+1\rangle$ and $|2,j\rangle$, so that
\begin{eqnarray}
|\psi_2\rangle &=& S(\theta) |\psi_1\rangle \nonumber \\
&=& \tilde{c}_j |2,j\rangle + \tilde{c}_{j+1} |1,j+1\rangle +  \sum_{n \ne j,j+1} c_n |0,n\rangle, \nonumber \\
\end{eqnarray}
where $\theta = \lambda g t$ and the new amplitudes are
\begin{equation}
\begin{array}{lcl}
 \tilde{c}_j &=& \cos \theta c_{j} - i \sin \theta c_{j+1} \\
  \tilde{c}_{j+1} &=& \cos \theta c_{j+1} - i \sin \theta c_j 
\end{array}
\end{equation}

To remove the entanglement between the qubit and the oscillator, we reverse the encoding step.  That is, we perform the number-state-dependent $\pi$-pulse $R_{01}^{(j+1)}$, the $q=2 \to 1$ transition $R_{12}$, and finally $R_{01}^{(j)}$.  The net result is to map $|2,j\rangle \to |0,j\rangle$ and $|1,j+1\rangle \to |0,j+1\rangle$, so that
\begin{eqnarray}
|\psi_3\rangle &=&  U_{\subs{decode}} |\psi_2\rangle \nonumber \\
&=& \tilde{c}_j |0,j\rangle + \tilde{c}_{j+1} |0,j+1\rangle +  \sum_{n \ne j,j+1} c_n |0,n\rangle, \nonumber \\
\end{eqnarray}
where $U_{\subs{decode}} = R_{01}^{(j)} R_{12} R_{01}^{(j+1)}$.   This has achieved the desired rotation, $\mathcal{R}_{j,j+1}(\theta)$.  In short, we have found
\begin{equation}
\mathcal{R}_{j,j+1}(\theta) = R_{01}^{(j)} R_{12} R_{01}^{(j+1)} S(\theta) R_{01}^{(j+1)} R_{12} R_{01}^{(j)}.
\end{equation}

As alluded to above, any Fock-state rotation $\mathcal{R}_{j,k}(\theta)$ can be implemented by using the nearest-neighbor rotations $\mathcal{R}_{j,j+1}(\theta)$.  This is done by swapping state amplitudes along paths in a ``coupling graph'', as described in \cite{Brennen05}.   For example, we can extend our construction to the rotations
\begin{eqnarray}
\mathcal{R}_{j,j+2}(\theta) &=& R_{01}^{(j)} R_{12} S(\pi) R_{12} R_{01}^{(j+1)} S(\theta)  \\ \nonumber
& & \times R_{01}^{(j+2)}  R_{12} S(\pi) R_{12} R_{01}^{(j)}.
\end{eqnarray}
and
\begin{eqnarray}
\mathcal{R}_{j,j+3}(\theta) &=& R_{01}^{(j)} R_{12} S(\pi) R_{12} S(\pi) \\ \nonumber
& & \times  R_{12} R_{01}^{(j+1)} S(\theta) R_{01}^{(j+2)} R_{12}  \\ \nonumber
& & \times  S(\pi) R_{12} S(\pi) R_{12} R_{01}^{(j)}.
\end{eqnarray}
Note that each of these has the form $U_{\subs{decode}} S(\theta) U_{\subs{encode}}$: we first transform the state by encoding it into a particular set of qudit states (suitably entangled with the auxiliary), perform a swap, and then decode the state so that the net result is a transformation of the qudit state alone.

\section{Numerical Simulation}

\begin{figure}
\begin{center}
\includegraphics[width=3in]{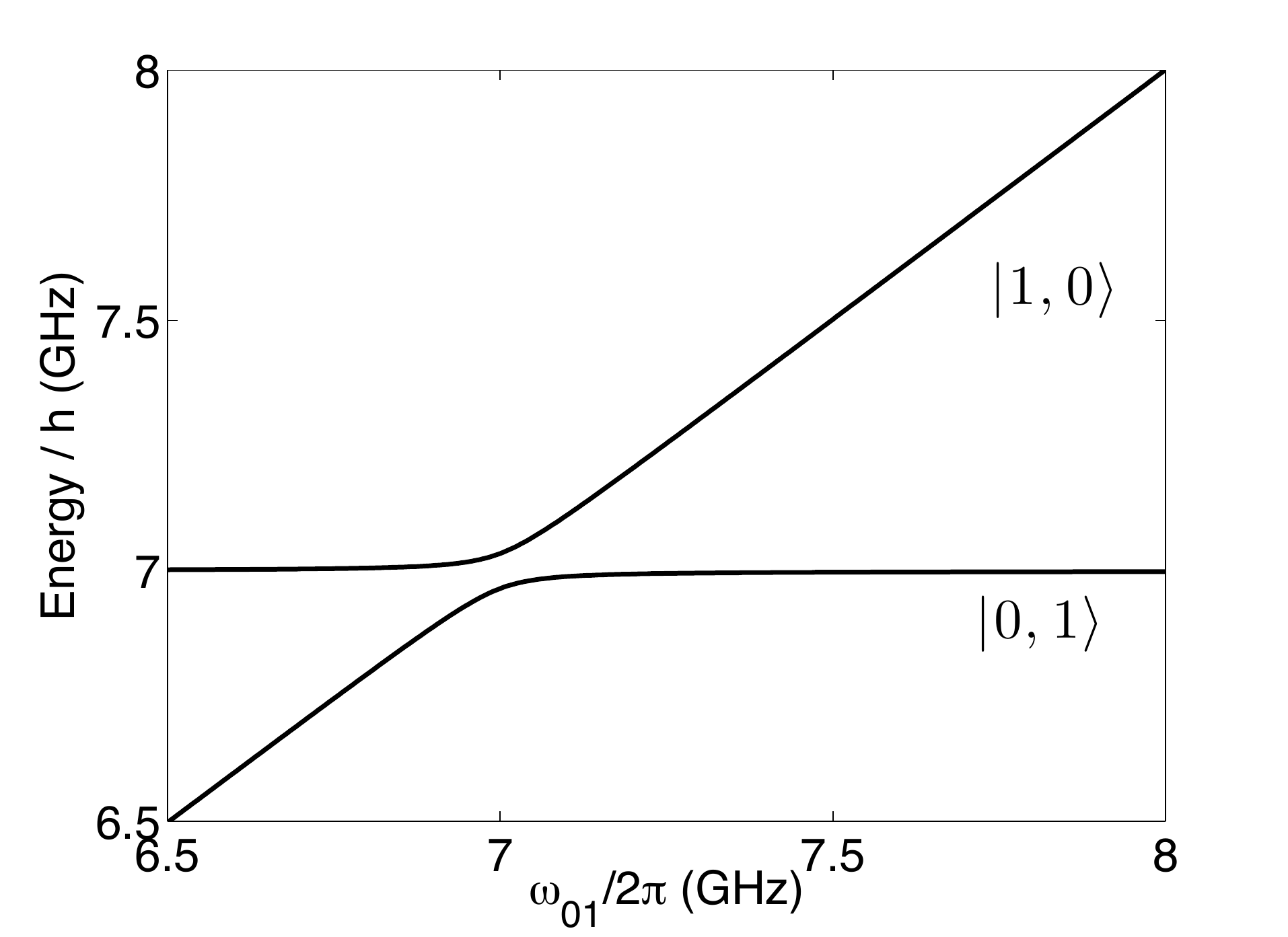}
\end{center}
\caption{ The energy levels $E_1$ and $E_2$ for the coupled qubit-resonator system, as a function of the qubit frequency $\omega_{01}$.  These levels correspond to the single-excitation subspace, with eigenstates $|\Psi_1\rangle \approx |0,1\rangle$ and $|\Psi_2\rangle \approx |1,0\rangle$ for $\omega_{01}/2\pi > 7.3 \ \mbox{GHz}$. }
\label{level1}
\end{figure}

\begin{figure}
\begin{center}
\includegraphics[width=3in]{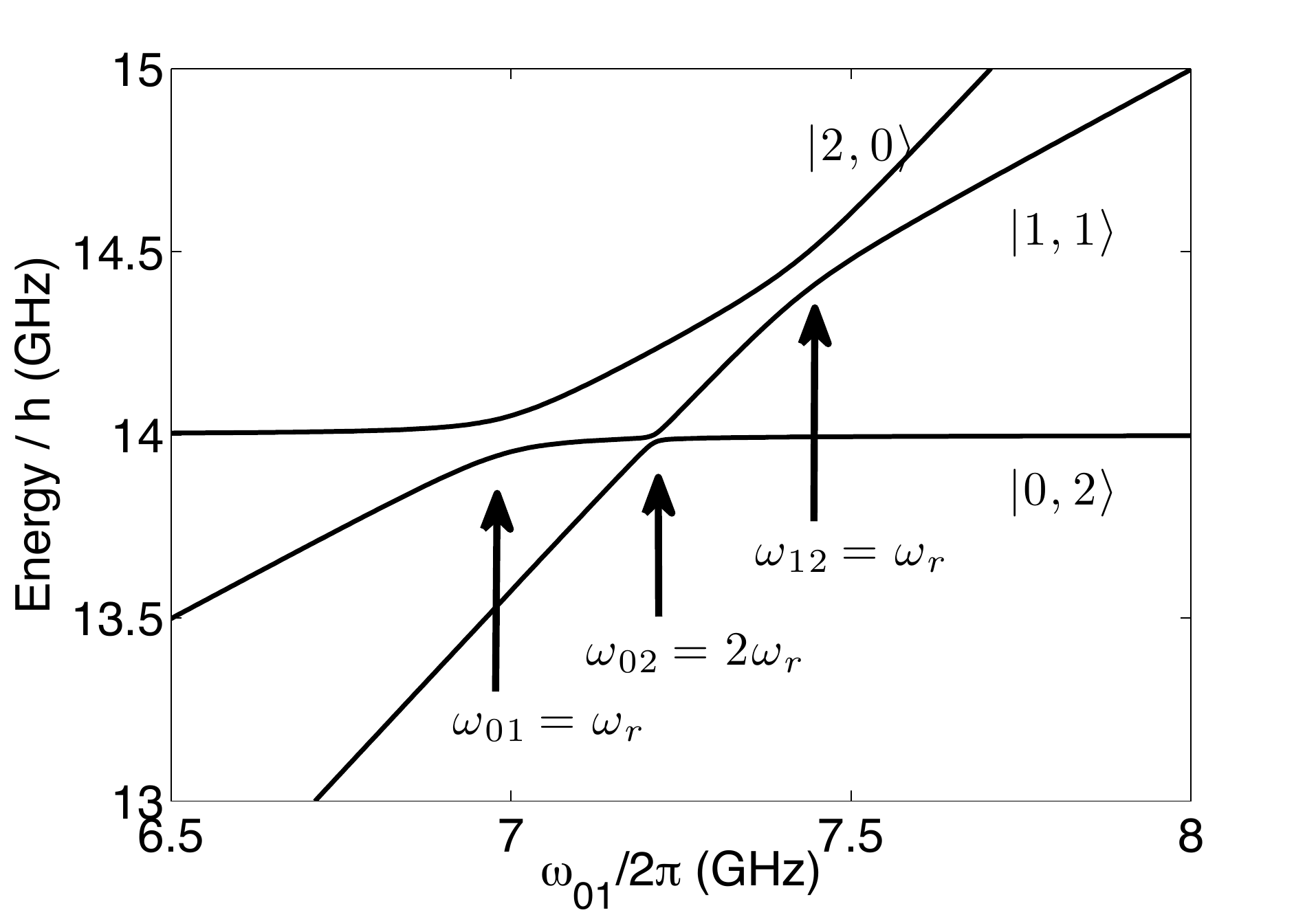}
\end{center}
\caption{ The energy levels $E_3$, $E_4$, and $E_5$ for the coupled qubit-resonator system, as a function of the qubit frequency $\omega_{01}$.  These levels correspond to the two-excitation subspace, with eigenstates $|\Psi_3\rangle \approx |0,2\rangle$, $|\Psi_4\rangle \approx |1,1\rangle$, and $|\Psi_5\rangle \approx |2,0\rangle$ for $\omega_{01}/2\pi > 7.6 \ \mbox{GHz}$.}
\label{level2}
\end{figure}

We have solved the Schr{\"o}dinger equation for a four-level system coupled to a resonator.  The lowest few energy levels $E_n$, $n=0,1,2,\cdots$, for this system are shown in Figs. \ref{level1} and \ref{level2}.   We have used four levels for the auxiliary and ten for the resonator, with parameters similar to transmon-style qubits: $(\omega_{01}-\omega_{12})/2\pi = 420 \ \mbox{MHz}$, $(\omega_{01}-\omega_{23})/ 2\pi = 910 \ \mbox{MHz}$, $\omega_r/2\pi = 7 \ \mbox{GHz}$, and $g/2\pi = 35 \ \mbox{MHz}$.  These are similar to recent experiments, and the resulting levels are very similar to the energy levels for coupled phase qubits \cite{Strauch2003,Johnson2003}.     

\begin{figure}
\begin{center}
\includegraphics[width=3in]{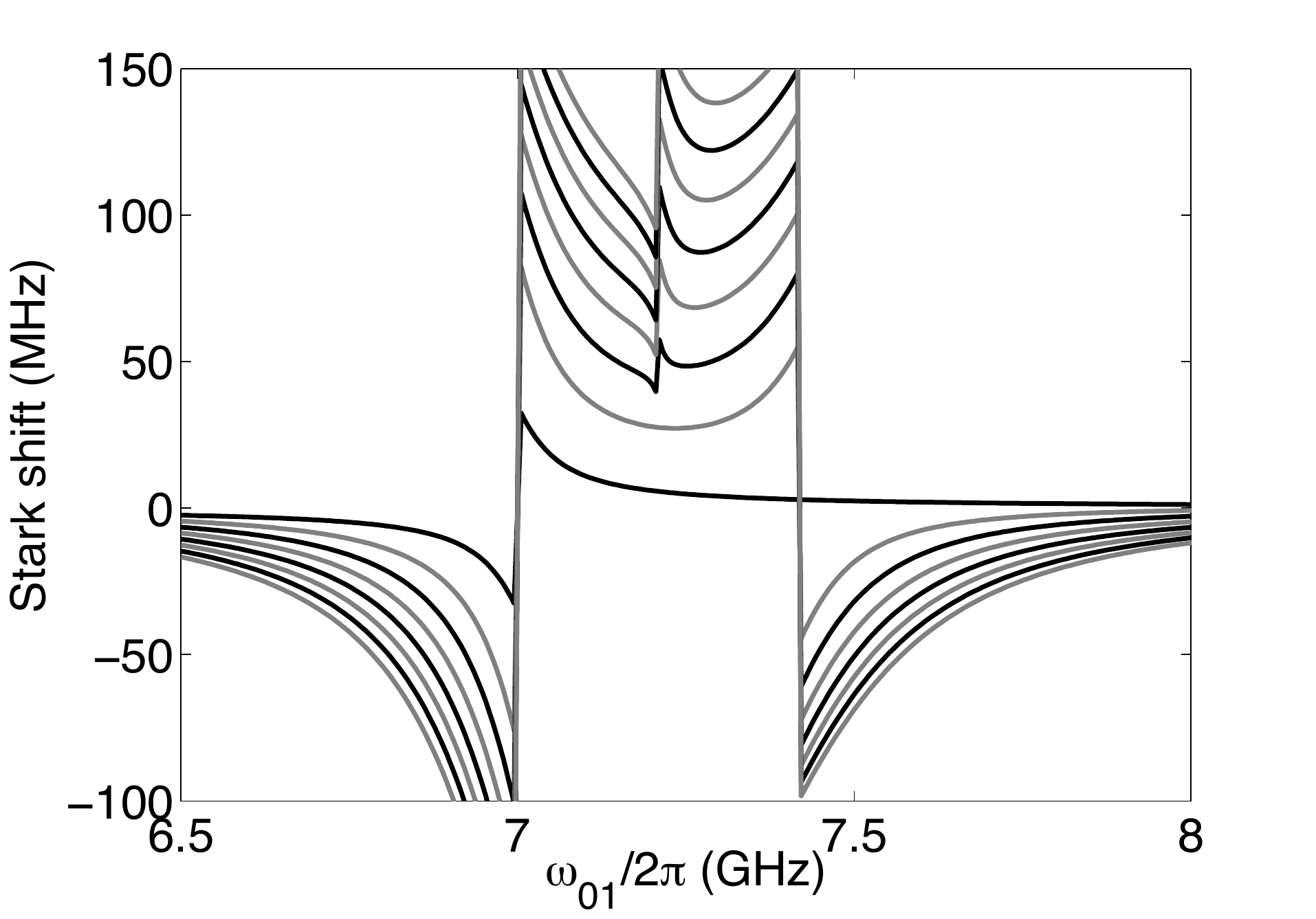}
\end{center}
\caption{ The number-state-dependent Stark shift $\omega_{01}^{(n)}-\omega_{01}$ as a function of the qubit frequency $\omega_{01}$.  The numerically calculated Stark shift is shown for $n=0 \to 7$.}
\label{level3}
\end{figure}

Three avoided crossings are indicated in Fig. \ref{level2}.  The first has $\omega_{01} = \omega_r$,  while the second, $\omega_{02} = 2 \omega_r$, is a second-order crossing.  The gate described above uses the third avoided crossing at $\omega_{12} = \omega_r$.   Away from these crossings, we can define a Stark shift for transitions between states predominantly composed of the uncoupled eigenstates $|q,n\rangle$.  In the following, we will use these ``dressed'' eigenstates to characterize our logic gate.  The Stark shift $\omega_{01}^{(n)} - \omega_{01}$ as a function of qubit frequency for the various Fock states is shown in Fig. \ref{level3}; the additional structure in the straddling regime is due to the second-order crossing.    

\begin{figure}
\begin{center}
\includegraphics[width=3in]{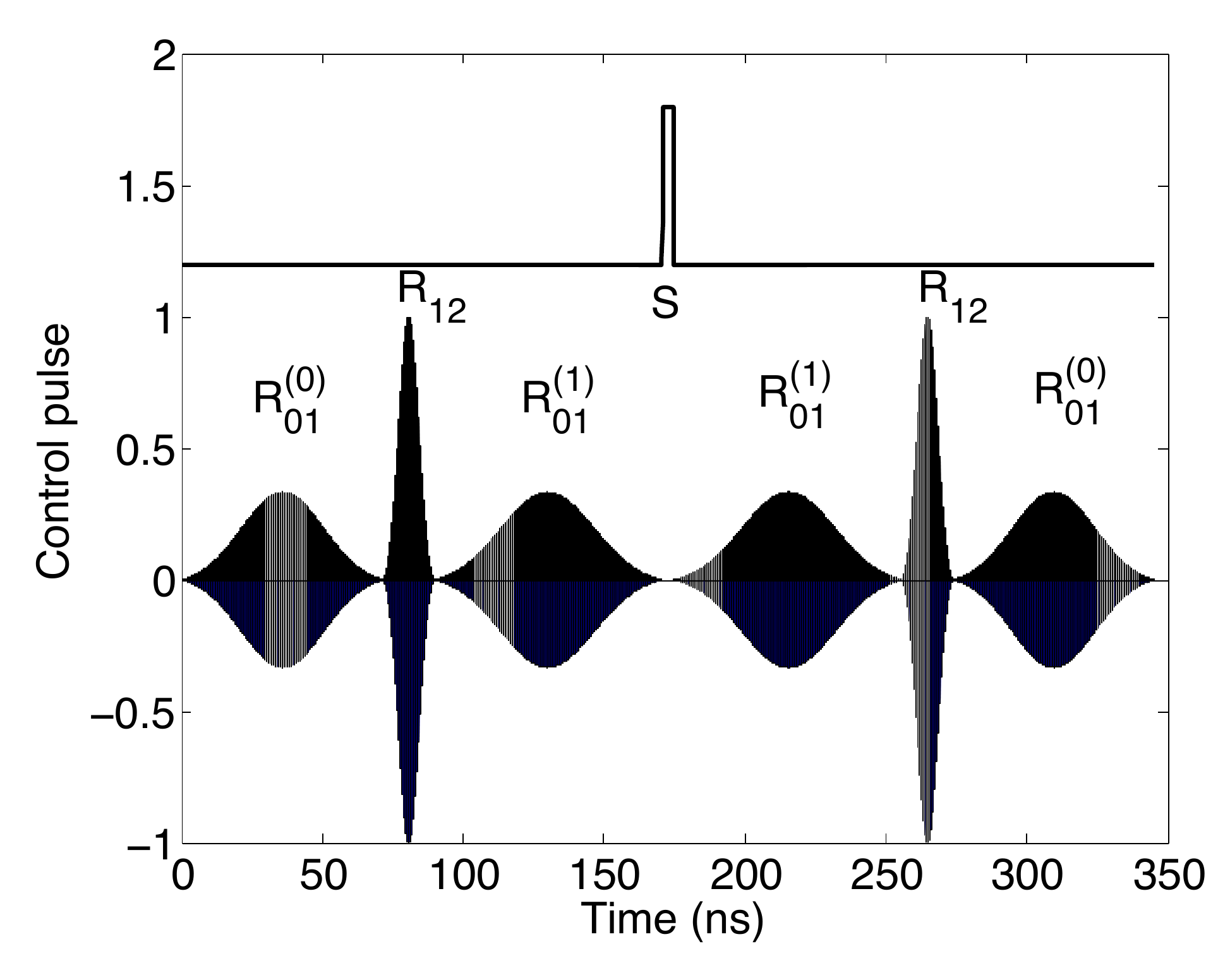}
\end{center}
\caption{ Control sequence for resonator logic gate $U_{0,1}$.  The upper line indicates the qubit frequency as a function of time, while the lower indicates the microwave pulses required to encode and decode the appropriate Fock states. The various steps of the sequence are labelled.}
\label{control}
\end{figure}

To illustrate the logic gate sequence described above, we start in the straddling regime with $\omega_{01}/2\pi = 7.28 \mbox{GHz}$, and implement the control sequence shown in Fig. \ref{control}.  The longer microwave pulses implement the number-state-dependent rotations, the shorter pulses the $q=1 \to 2$ transition, while the upper shift pulse implements the swap operation.  All of the microwave pulses use a truncated Gaussian profile \cite{Motzoi09}.  Here the qubit frequency is shifted from $\omega_{01}/2\pi = 7.28 \to 7.49 \ \mbox{GHz}$ and back, causing the exchange $|1,1\rangle \to |0,2\rangle$.  The amplitude or the timing of this shift pulse can be adjusted for an arbitrary rotation; here we have chosen to perform a full swap $n=0 \to 1$ or $n=1 \to 0$.  For this choice of system parameters, the complete sequence takes $346 \mbox{ns}$.  In terms of the dressed eigenstates, this swap is between the energy eigenstates $|\Psi_1\rangle$ and $|\Psi_2\rangle$.  This sequence can be extended to perform swaps between any neighboring Fock states with a similar control pulse.

\begin{figure}
\begin{center}
\includegraphics[width=3in]{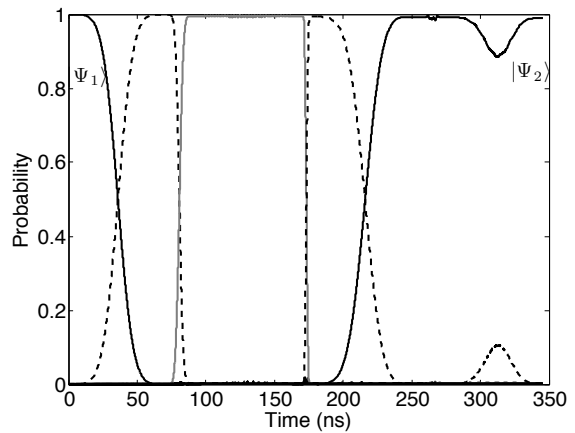}
\end{center}
\caption{ Time-dependent probabilities $p_k = |\langle \Psi_k |\Psi(t)\rangle|^2$ for the swap $n=0 \to 1$.  The solid black lines are for $k=1$ and 2, while the dashed and gray lines are for $k=3 \to 6$. In this simulation, the initial state $|\Psi(0)\rangle = |\Psi_1\rangle$ is evolved using the control sequence of Fig. \ref{control}.    }
\label{gate01}
\end{figure}

\begin{figure}
\begin{center}
\includegraphics[width=3in]{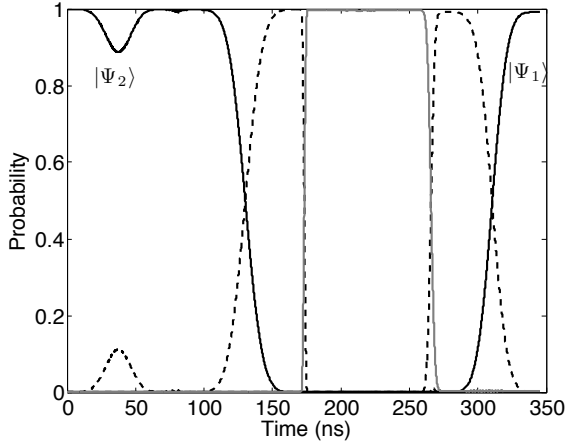}
\end{center}
\caption{ Time-dependent probabilities $p_k = |\langle \Psi_k |\Psi(t)\rangle|^2$ for the swap $n=1 \to 0$.  The solid black lines are for $k=1$ and 2, while the dashed and gray lines are for $k=3 \to 6$. In this simulation, the initial state $|\Psi(0)\rangle = |\Psi_2\rangle$ is evolved using the control sequence of Fig. \ref{control}.    }
\label{gate10}
\end{figure}

Solving the time-dependent Schr{\"o}dinger equation, the probabilities $p_k = |\langle \Psi_k |\Psi(t)\rangle|^2$ (for the first few eigenstates) are shown in Figs. \ref{gate01} and \ref{gate10}, using initial conditions appropriate to $n=0$ and $n=1$, respectively.  The swap probability for these states is $\sim 0.99$ , while the higher Fock states are unaffected (with fidelity $> 0.95$).  

\section{Two-Qudit Gate}

An extension of this scheme to two-qudit, and hence arbitrary quantum computation, will now be described. To perform the two-qudit operation $\mathcal{U}_{j,k}$, we return to the circuit of Fig. \ref{fig2}, with an auxiliary qutrit for each qudit, and the qudits are coupled together.  We denote the states by $|q_a, q_b, n_a, n_b\rangle$.  We again use the number-state-dependent rotations to encode and decode the Fock state $|j,k\rangle$ to be coupled.  Starting with
\begin{equation}
|\psi_0\rangle = |0,0\rangle \otimes \sum_{n,m} c_{n,m} |n,m\rangle,
\end{equation}
the encoding operation $U_{\subs{encode}} = R_{A,01}^{(j)} R_{B,01}^{(k)}$ prepares the system in the state
\begin{equation}
|\psi_1\rangle = U_{\subs{encode}} |\psi_0\rangle =  c_{j,k} |1,1,j,k\rangle  + |\delta \psi\rangle,
\end{equation}
where
\begin{eqnarray}
|\delta \psi\rangle &=& \sum_{m \ne k} c_{j,m} |1,0,j,m\rangle + \sum_{n \ne j} c_{n,k} |0,1,n,k\rangle \nonumber \\
& & + \sum_{n \ne j, m\ne k} c_{n,m} |0,0,n,m\rangle.
\end{eqnarray}
This operation has selected out the oscillator states with $n_a = j$ and $n_b = k$.

A quantum logic operation can now be performed on the qutrits, specifically a controlled-phase gate $C(\theta)$ of the form $|q_a,q_b\rangle \to e^{i \phi_{ab}} |q_a,q_b\rangle$, where $\phi_{ab} = \theta$ for $q_a=q_b = 1$ and zero otherwise.  This generates the transformation
\begin{equation}
|\psi_2\rangle = C(\theta) |\psi_1\rangle = e^{i \theta} c_{j,k} |1,1, j,k\rangle + |\delta \psi\rangle.
\end{equation}
Finally, by using $U_{\subs{decode}} = R_{A,01}^{(j)} R_{B,01}^{(k)}$, we find
\begin{eqnarray}  
|\psi_3\rangle &=& U_{\subs{decode}} |\psi_2\rangle \nonumber \\
&=& e^{i \theta} c_{j,k} |0,0,j,k\rangle + \sum_{n,m \ne (j,k)} c_{n,m} |0,0,n,m\rangle, \nonumber \\
\end{eqnarray}
returning the encoded states to the resonator.  In summary, we have shown that
\begin{equation}
\mathcal{U}_{j,k}(\theta) = R_{A,01}^{(j)} R_{B,01}^{(k)} C(\theta) R_{A,01}^{(j)} R_{B,01}^{(k)}.
\end{equation}
Combining logic gates of this form with single-qudit operations allows for universal quantum computation over an arbitrary number of qudits \cite{Brennen05}

\begin{figure}
\begin{center}
\includegraphics[width=3in]{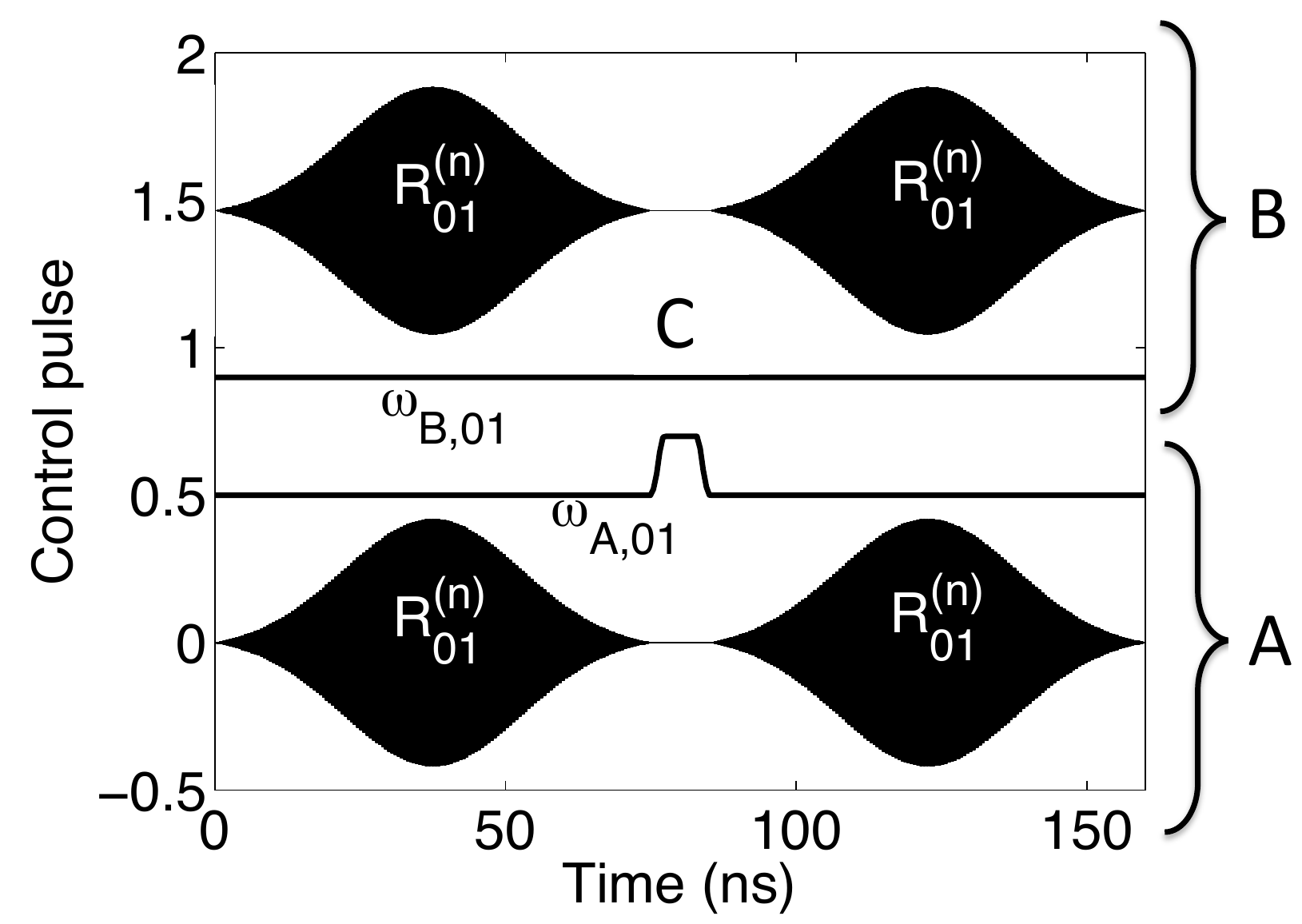}
\end{center}
\caption{Control sequence for resonator two-qudit logic gate $\mathcal{U}_{n,n}$.  The upper two curves indicates the microwave pulses and qubit frequency as a function of time for system $B$, while the lower two indicate the qubit frequency for system $B$.  The various steps of the sequence are labelled, with $C$ corresponding the controlled-phase gate operated at $\omega_{A,01} = \omega_{B,12}$ (see text). }
\label{cphase}
\end{figure}

The controlled-phase gate $C(\theta)$ between the two qutrits can be implemented by shifting their frequencies so that $\omega_{A,01} = \omega_{B,12}$.  The interaction $\hbar g_{ab} (\sigma_{+}^{A} \sigma_{-}^{B}~+~\sigma_{-}^{A} \sigma_{+}^{B} )$ now leads to the resonant exchange $|1,1\rangle \to -i |0,2\rangle \to - |1,1\rangle$ \cite{Strauch2003}.  This $\pi$-phase shift can be adjusted to any value by using a nonzero detuning $\omega_{A,01} - \omega_{B,12}$ \cite{Mariantoni11}, or by an adiabatic implementation \cite{DiCarlo2009}.  The full control pulse, assuming $g_{ab}/2\pi = 35 \mbox{MHz}$, is shown in Fig. \ref{cphase}, taking a total time of 150 ns.  

\section{Decoherence and Measurement}

Resonators have very attractive coherence properties, with the potential for more complex qudit operations than their qubit counterparts.  Resonators have shown nearly ideal decoherence dynamics \cite{Wang2008}, described mainly by energy loss.  On-chip resonators typically have coherence times greater than $1 \ \mu \mbox{s}$, while recent three-dimensional cavities have shown qubit coherence times $T_q$ greater than $10 \ \mu \mbox{s}$ and resonator coherence times $T_r$ greater than $50 \ \mu \mbox{s}$ \cite{Paik2011}.  However, the $n$-th excited state of the resonator decays with a rate $n / T_r$, proportional to the Fock state number.  A reasonable conclusion is that a ``good'' resonator qudit could have $d < 1 + T_r/T_q \approx 6-10$.   We will numerically simulate the gate sequences described above to verify this conclusion.

Resonator qudits will also require require a means to readout the resonator state.  The simplest method would use the quantum Rabi oscillations for $\omega_{01} = \omega_r$, as in the experiments of Hofheinz {\it et al.} \cite{Hofheinz2008,Hofheinz2009}.  Here the exchange of energy between qubit-resonator states $|0,n\rangle$ and $|1,n-1\rangle$ occurs with (angular) frequency $g \sqrt{n}$.  This allows the populations of the various Fock states to be found by collecting a suitably long time-series and Fourier analysis.  An alternative method would use the number-state-dependent Rabi transitions to implement the non-demolition method of Johnson {\it et al.} \cite{Johnson10}.   A sequence of such transitions applied to a qubit initially in its ground state would allow the populations of the various Fock states to be determined, one by one.  Both methods would require repeated qubit measurements to estimate the Fock state probabilities.  

\subsection{Decoherence Simulation}

We model decoherence using the Lindblad master equation
\begin{equation}
\frac{d \rho}{dt} = - \frac{i}{\hbar} [\mathcal{H}, \rho] + \sum_{j} \lambda_j \left(L_j \rho L_j^{\dagger} - \frac{1}{2} L_j^{\dagger} L_j \rho - \frac{1}{2} \rho L_j^{\dagger} L_j \right),
\end{equation} 
with up to four Lindblad operators $L_1 = \sigma_-^{A}$, $L_2 = \sigma_-^{B}$, $L_3 = a$, $L_4 = b$, and rates $\lambda_1 = \lambda_2 = 1/T_q$ and $\lambda_3 = \lambda_4 = 1/T_r$.  To simplify the calculation, we transform to an interaction picture and keep only the resonant terms in $\mathcal{H}$ for each step of the logic gate.  

\begin{figure}
\begin{center}
\includegraphics[width=3in]{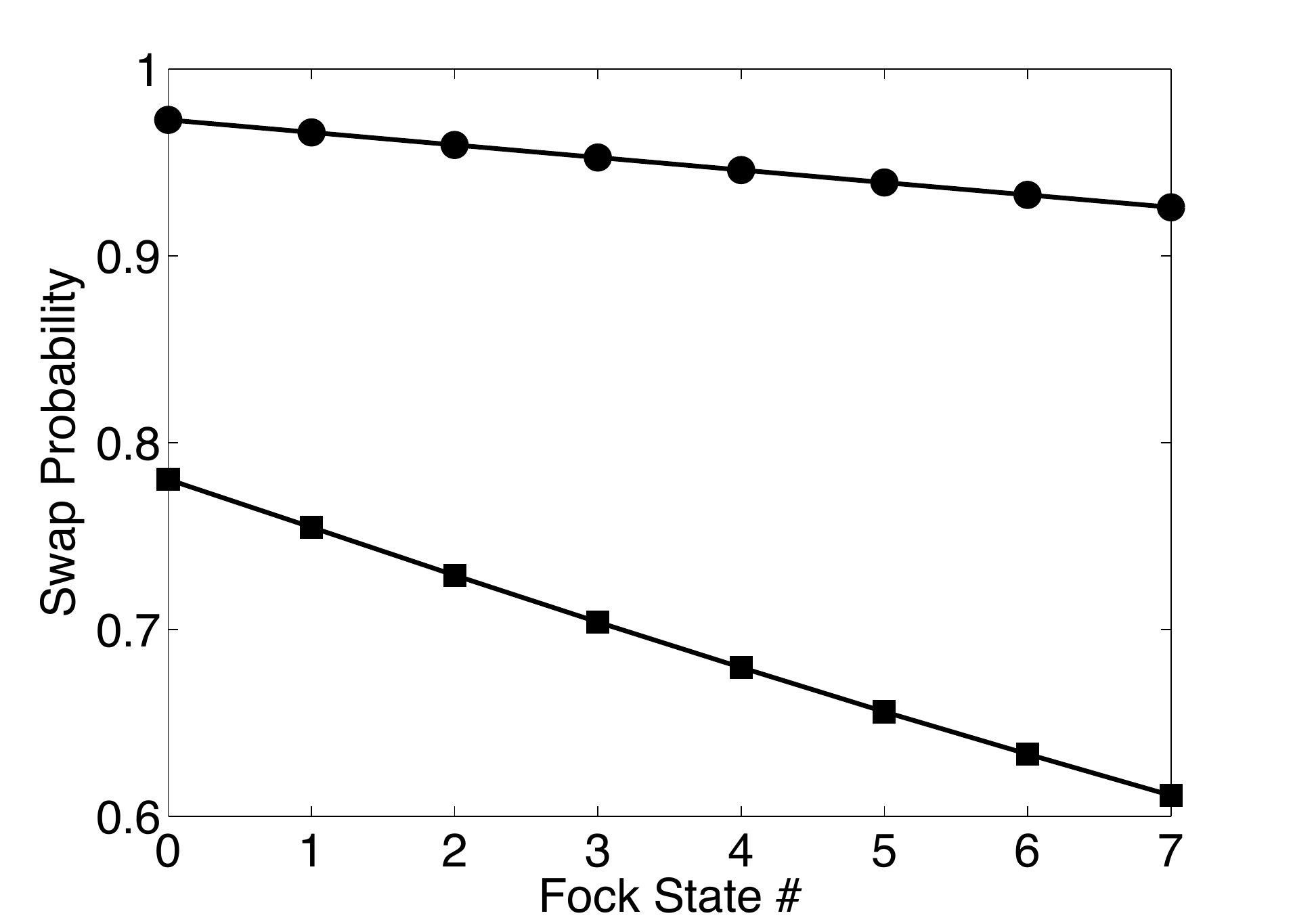}
\end{center}
\caption{ Swap probabilities for the single-qudit rotations $U_{n,n+1}(\pi)$, for $n=0 \to 7$.  The upper dots are for coherence times $T_q = 10 \ \mu \mbox{s}, T_r = 50 \ \mu \mbox{s}$, while the lower squares are for coherence times $T_q = 1 \ \mu \mbox{s}, T_r = 10  \ \mu \mbox{s}$.  Other relevant parameters are $\Omega_1/2\pi = 6.67 \mbox{MHz}$, $\Omega_2/2\pi = 25 \mbox{MHz}$, and $g/2\pi = 35 \mbox{MHz}$.  The curves are guides for the eye.}
\label{decgate1}
\end{figure}

For the single-qudit logic gate, this entails the sequence of interactions 
\begin{equation}
\begin{array}{lcll}
\mathcal{H}_{\mbox{\scriptsize{int}}}/\hbar &=& \frac{1}{2} \Omega_1 (|0\rangle \langle 1| + |1\rangle \langle 0|) \otimes |j\rangle \langle j| & \mbox{for} \ R_{01}^{(j)}, \\
&=& \frac{1}{2} \Omega_2 (|1\rangle \langle 2| + |2\rangle \langle 1|) \otimes I & \mbox{for} \ R_{12}, \\
&=& g (a \sigma_+ + a^{\dagger} \sigma_) & \mbox{for} \ S(\theta),
\end{array}
\end{equation}
where $I$ is the identity operator for the resonator.  The resulting swap probabilities for the single-qudit rotations $U_{n,n+1}(\pi)$ are shown in Fig. \ref{decgate1}, for Fock states $n = 0 \to 7$.  These simulations use a quantum trajectories approach to integrate the master equation,with 1024 trajectories.  The upper curve is for state-of-the-art coherence times, while the lower is for typical on-chip circuits.  These results are consistent with a loss of coherence proportional to $e^{-0.66 T/T_q} e^{-nT/T_r}$, where $T = 342-346 \mbox{ns}$ is the total time for the single-qudit gate.  As discussed above, the resonator Fock states with $n < T_r/T_q$ will have errors of about the same order as a single-qubit gate of the same duration.  

\begin{figure}
\begin{center}
\includegraphics[width=3in]{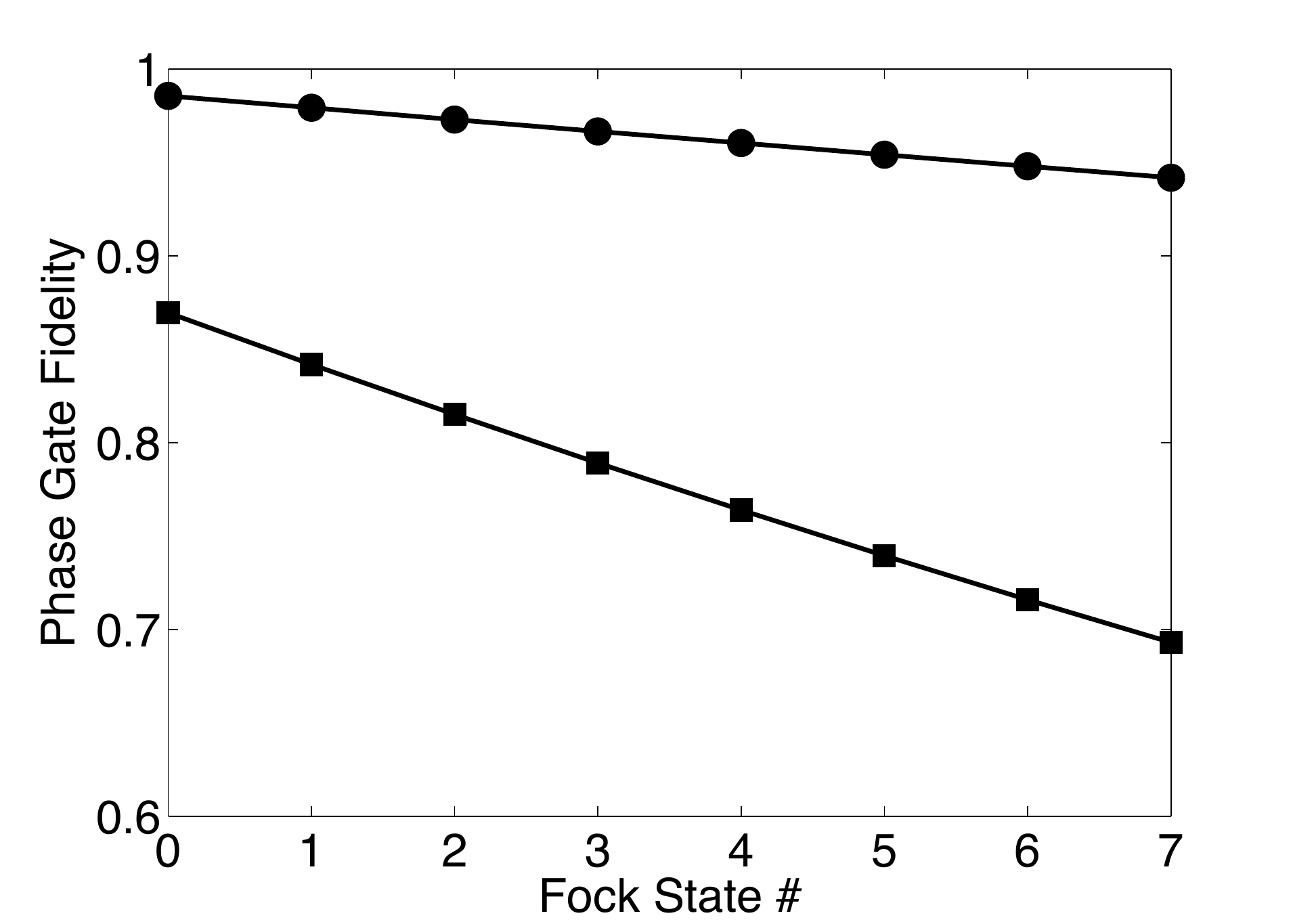}
\end{center}
\caption{ Gate fidelities for the two-qudit phase gate $\mathcal{U}_{n,n}(\pi)$ for $n=0 \to 7$.  The upper dots are for coherence times $T_q = 10 \ \mu \mbox{s}, T_r = 50 \ \mu \mbox{s}$, while the lower squares are for coherence times $T_q = 1 \ \mu \mbox{s}, T_r = 10 \ \mu \mbox{s}$. Other relevant parameters are $\Omega/2\pi = 6.67 \mbox{MHz}$ and $g_{ab}/2\pi = 35 \mbox{MHz}$.   The curves are guides for the eye.}
\label{decgate2}
\end{figure}

For the two-qudit controlled-phase gate, the interaction Hamiltonians are
\begin{equation}
\begin{array}{lcll}
\mathcal{H}_{\mbox{\scriptsize{int}}}/\hbar &=& \frac{1}{2} \Omega \sigma_x \otimes I_B \otimes |j\rangle \langle j| \otimes I & \mbox{for} \ R_{A,01}^{(j)}, \\
&=& \frac{1}{2} \Omega I_A \otimes \sigma_x \otimes I \otimes |j\rangle \langle j| & \mbox{for} \ R_{B,01}^{(j)}, \\
&=& \sqrt{2} g \left(|11\rangle \langle 02| + |02\rangle \langle 11|\right) \otimes I \otimes I & \mbox{for} \ C(\theta),
\end{array}
\end{equation}
where $I_A$, $I_B$, and $I$ are the identity operators for the auxiliaries $A$, $B$, and a resonator, respectively.  The resulting (worst-case) fidelities for the two-qudit gate $\mathcal{U}_{n,n}(\pi)$ are shown in Fig. \ref{decgate2}, again calculated using the quantum trajectories method for the master equation.  These results are somewhat better than the single-qudit gate, proportional to $e^{-T/T_q} e^{-2 n T/T_r}$, here with a smaller overall time $T = 160 \ \mbox{ns}$.  

\section{Conclusion}

We have presented an approach to quantum computation using the multilevel Hilbert space of a resonator as a new type of qudit.  This approach is based on resonant and dispersive interactions that have been demonstrated experimentally, and can be extended to multi-resonator logic gates.  Thus, a successful demonstration of this scheme will open up a number of interesting questions in quantum information processing.  

First, while there is a great deal known about the theory of quantum circuits and algorithms for qubits \cite{mikeandike}, much remains to be learned about qudit algorithms.  While the asymptotic complexity should be identical \cite{Bullock05}, these results indicate that an arbitrary unitary gate on $n$ qudits requires $d^n$ elementary gates (exponential in $n$).   Efficient quantum algorithms, using specific gates such as the quantum Fourier transform, can be implemented using a polynomial number of qubit gates.  An interesting problem would be to determine if qudit constructions for the Fourier transform can be more efficient than the qubit constructions.

Second, our analytical approach leaves open questions about optimization of operations in this larger Hilbert space.  There may be interesting approaches to construct a given unitary of interest that is more efficient than the two-level reduction used here.  In particular, the number-state-dependent transitions dominate the operation time.  While this time can likely be decreased by using larger couplings or more sophisticated microwave pulses, perhaps using multiple frequencies \cite{Forney10} or multiple quadratures \cite{Motzoi09}, other approaches may be necessary.  Optimal control methods \cite{Merkel09} for this system may lead to such alternative approaches, and would be important for operations in the presence of decoherence.   

Finally, the two measurement approaches presented both use qubit measurements to read out the resonator states. Either approach should allow for full tomography of the resonator logic gates, at the expense of having to perform a large number of experiments to determine the state of the resonator.  An open question is how to extract this information in the most direct and efficient manner.  These and other issues will be fruitful tests of our understanding of quantum control and measurement of multilevel quantum systems.

\acknowledgments
I gratefully acknowledge discussions with K. Jacobs, B. Johnson, and R. W. Simmonds.  This work was supported by NSF grant PHY-1005571.

\bibliography{report} 

\begin{thebibliography}{48}
\expandafter\ifx\csname natexlab\endcsname\relax\def\natexlab#1{#1}\fi
\expandafter\ifx\csname bibnamefont\endcsname\relax
  \def\bibnamefont#1{#1}\fi
\expandafter\ifx\csname bibfnamefont\endcsname\relax
  \def\bibfnamefont#1{#1}\fi
\expandafter\ifx\csname citenamefont\endcsname\relax
  \def\citenamefont#1{#1}\fi
\expandafter\ifx\csname url\endcsname\relax
  \def\url#1{\texttt{#1}}\fi
\expandafter\ifx\csname urlprefix\endcsname\relax\def\urlprefix{URL }\fi
\providecommand{\bibinfo}[2]{#2}
\providecommand{\eprint}[2][]{\url{#2}}

\bibitem[{\citenamefont{Clarke and Wilhelm}(2008)}]{Clarke2008}
\bibinfo{author}{\bibfnamefont{J.}~\bibnamefont{Clarke}} \bibnamefont{and}
  \bibinfo{author}{\bibfnamefont{F.~K.} \bibnamefont{Wilhelm}},
  \bibinfo{journal}{Nature} \textbf{\bibinfo{volume}{453}},
  \bibinfo{pages}{1031} (\bibinfo{year}{2008}).

\bibitem[{\citenamefont{DiCarlo et~al.}(2009)\citenamefont{DiCarlo, Chow,
  Gambetta, Bishop, Johnson, Schuster, Majer, Blais, Frunzio, Girvin
  et~al.}}]{DiCarlo2009}
\bibinfo{author}{\bibfnamefont{L.}~\bibnamefont{DiCarlo}},
  \bibinfo{author}{\bibfnamefont{J.~M.} \bibnamefont{Chow}},
  \bibinfo{author}{\bibfnamefont{J.~M.} \bibnamefont{Gambetta}},
  \bibinfo{author}{\bibfnamefont{L.~S.} \bibnamefont{Bishop}},
  \bibinfo{author}{\bibfnamefont{B.~R.} \bibnamefont{Johnson}},
  \bibinfo{author}{\bibfnamefont{D.~I.} \bibnamefont{Schuster}},
  \bibinfo{author}{\bibfnamefont{J.}~\bibnamefont{Majer}},
  \bibinfo{author}{\bibfnamefont{A.}~\bibnamefont{Blais}},
  \bibinfo{author}{\bibfnamefont{L.}~\bibnamefont{Frunzio}},
  \bibinfo{author}{\bibfnamefont{S.~M.} \bibnamefont{Girvin}},
  \bibnamefont{et~al.}, \bibinfo{journal}{Nature}
  \textbf{\bibinfo{volume}{460}}, \bibinfo{pages}{240} (\bibinfo{year}{2009}).

\bibitem[{\citenamefont{DiCarlo et~al.}(2010)\citenamefont{DiCarlo, Reed, Sun,
  Johnson, Chow, Gambetta, Frunzio, Girvin, Devoret, and
  Schoelkopf}}]{DiCarlo2010}
\bibinfo{author}{\bibfnamefont{L.}~\bibnamefont{DiCarlo}},
  \bibinfo{author}{\bibfnamefont{M.~D.} \bibnamefont{Reed}},
  \bibinfo{author}{\bibfnamefont{L.}~\bibnamefont{Sun}},
  \bibinfo{author}{\bibfnamefont{B.~R.} \bibnamefont{Johnson}},
  \bibinfo{author}{\bibfnamefont{J.~M.} \bibnamefont{Chow}},
  \bibinfo{author}{\bibfnamefont{J.~M.} \bibnamefont{Gambetta}},
  \bibinfo{author}{\bibfnamefont{L.}~\bibnamefont{Frunzio}},
  \bibinfo{author}{\bibfnamefont{S.~M.} \bibnamefont{Girvin}},
  \bibinfo{author}{\bibfnamefont{M.~H.} \bibnamefont{Devoret}},
  \bibnamefont{and} \bibinfo{author}{\bibfnamefont{R.~J.}
  \bibnamefont{Schoelkopf}}, \bibinfo{journal}{Nature}
  \textbf{\bibinfo{volume}{467}}, \bibinfo{pages}{574} (\bibinfo{year}{2010}).

\bibitem[{\citenamefont{Yamamoto et~al.}(2010)\citenamefont{Yamamoto, Neeley,
  Lucero, Bialczak, Kelly, Lenander, Mariantoni, O'Connell, Sank, Wang
  et~al.}}]{Yamamoto10}
\bibinfo{author}{\bibfnamefont{T.}~\bibnamefont{Yamamoto}},
  \bibinfo{author}{\bibfnamefont{M.}~\bibnamefont{Neeley}},
  \bibinfo{author}{\bibfnamefont{E.}~\bibnamefont{Lucero}},
  \bibinfo{author}{\bibfnamefont{R.~C.} \bibnamefont{Bialczak}},
  \bibinfo{author}{\bibfnamefont{J.}~\bibnamefont{Kelly}},
  \bibinfo{author}{\bibfnamefont{M.}~\bibnamefont{Lenander}},
  \bibinfo{author}{\bibfnamefont{M.}~\bibnamefont{Mariantoni}},
  \bibinfo{author}{\bibfnamefont{A.~D.} \bibnamefont{O'Connell}},
  \bibinfo{author}{\bibfnamefont{D.}~\bibnamefont{Sank}},
  \bibinfo{author}{\bibfnamefont{H.}~\bibnamefont{Wang}}, \bibnamefont{et~al.},
  \bibinfo{journal}{Phys. Rev. B} \textbf{\bibinfo{volume}{82}},
  \bibinfo{pages}{184515} (\bibinfo{year}{2010}).

\bibitem[{\citenamefont{Strauch et~al.}(2003)\citenamefont{Strauch, Johnson,
  Dragt, Lobb, Anderson, and Wellstood}}]{Strauch2003}
\bibinfo{author}{\bibfnamefont{F.~W.} \bibnamefont{Strauch}},
  \bibinfo{author}{\bibfnamefont{P.~R.} \bibnamefont{Johnson}},
  \bibinfo{author}{\bibfnamefont{A.~J.} \bibnamefont{Dragt}},
  \bibinfo{author}{\bibfnamefont{C.~J.} \bibnamefont{Lobb}},
  \bibinfo{author}{\bibfnamefont{J.~R.} \bibnamefont{Anderson}},
  \bibnamefont{and} \bibinfo{author}{\bibfnamefont{F.~C.}
  \bibnamefont{Wellstood}}, \bibinfo{journal}{Phys. Rev. Lett.}
  \textbf{\bibinfo{volume}{91}}, \bibinfo{pages}{167005}
  (\bibinfo{year}{2003}).

\bibitem[{\citenamefont{Lanyon et~al.}(2009)\citenamefont{Lanyon, Barbieri,
  Almeida, Jennewein, Ralph, resch, Pryde, O'Brien, Gilchrist, and
  White}}]{Lanyon09}
\bibinfo{author}{\bibfnamefont{B.~P.} \bibnamefont{Lanyon}},
  \bibinfo{author}{\bibfnamefont{M.}~\bibnamefont{Barbieri}},
  \bibinfo{author}{\bibfnamefont{M.~P.} \bibnamefont{Almeida}},
  \bibinfo{author}{\bibfnamefont{T.}~\bibnamefont{Jennewein}},
  \bibinfo{author}{\bibfnamefont{T.~C.} \bibnamefont{Ralph}},
  \bibinfo{author}{\bibfnamefont{K.~J.} \bibnamefont{resch}},
  \bibinfo{author}{\bibfnamefont{G.~J.} \bibnamefont{Pryde}},
  \bibinfo{author}{\bibfnamefont{J.~L.} \bibnamefont{O'Brien}},
  \bibinfo{author}{\bibfnamefont{A.}~\bibnamefont{Gilchrist}},
  \bibnamefont{and} \bibinfo{author}{\bibfnamefont{A.~G.} \bibnamefont{White}},
  \bibinfo{journal}{Nature Physics} \textbf{\bibinfo{volume}{5}},
  \bibinfo{pages}{134} (\bibinfo{year}{2009}).

\bibitem[{\citenamefont{Claudon et~al.}(2004)\citenamefont{Claudon, Balestro,
  Hekking, and Buisson}}]{Claudon2004}
\bibinfo{author}{\bibfnamefont{J.}~\bibnamefont{Claudon}},
  \bibinfo{author}{\bibfnamefont{F.}~\bibnamefont{Balestro}},
  \bibinfo{author}{\bibfnamefont{F.~W.~J.} \bibnamefont{Hekking}},
  \bibnamefont{and} \bibinfo{author}{\bibfnamefont{O.}~\bibnamefont{Buisson}},
  \bibinfo{journal}{Phys. Rev. Lett.} \textbf{\bibinfo{volume}{93}},
  \bibinfo{pages}{187003} (\bibinfo{year}{2004}).

\bibitem[{\citenamefont{Dutta et~al.}(2008)\citenamefont{Dutta, Strauch, Lewis,
  Mitra, Paik, Palomaki, Tiesinga, Anderson, Dragt, Lobb et~al.}}]{Dutta2008}
\bibinfo{author}{\bibfnamefont{S.~K.} \bibnamefont{Dutta}},
  \bibinfo{author}{\bibfnamefont{F.~W.} \bibnamefont{Strauch}},
  \bibinfo{author}{\bibfnamefont{R.~M.} \bibnamefont{Lewis}},
  \bibinfo{author}{\bibfnamefont{K.}~\bibnamefont{Mitra}},
  \bibinfo{author}{\bibfnamefont{H.}~\bibnamefont{Paik}},
  \bibinfo{author}{\bibfnamefont{T.~A.} \bibnamefont{Palomaki}},
  \bibinfo{author}{\bibfnamefont{E.}~\bibnamefont{Tiesinga}},
  \bibinfo{author}{\bibfnamefont{J.~R.} \bibnamefont{Anderson}},
  \bibinfo{author}{\bibfnamefont{A.~J.} \bibnamefont{Dragt}},
  \bibinfo{author}{\bibfnamefont{C.~J.} \bibnamefont{Lobb}},
  \bibnamefont{et~al.}, \bibinfo{journal}{Phys. Rev. B}
  \textbf{\bibinfo{volume}{78}}, \bibinfo{pages}{104510}
  (\bibinfo{year}{2008}).

\bibitem[{\citenamefont{Neeley et~al.}(2009)\citenamefont{Neeley, Ansmann,
  Bialczak, Hofheinz, Lucero, O'Connell, Sank, Wang, Wenner, Cleland
  et~al.}}]{Neeley2009}
\bibinfo{author}{\bibfnamefont{M.}~\bibnamefont{Neeley}},
  \bibinfo{author}{\bibfnamefont{M.}~\bibnamefont{Ansmann}},
  \bibinfo{author}{\bibfnamefont{R.~C.} \bibnamefont{Bialczak}},
  \bibinfo{author}{\bibfnamefont{M.}~\bibnamefont{Hofheinz}},
  \bibinfo{author}{\bibfnamefont{E.}~\bibnamefont{Lucero}},
  \bibinfo{author}{\bibfnamefont{A.~D.} \bibnamefont{O'Connell}},
  \bibinfo{author}{\bibfnamefont{D.}~\bibnamefont{Sank}},
  \bibinfo{author}{\bibfnamefont{H.}~\bibnamefont{Wang}},
  \bibinfo{author}{\bibfnamefont{J.}~\bibnamefont{Wenner}},
  \bibinfo{author}{\bibfnamefont{A.~N.} \bibnamefont{Cleland}},
  \bibnamefont{et~al.}, \bibinfo{journal}{Science}
  \textbf{\bibinfo{volume}{325}}, \bibinfo{pages}{722} (\bibinfo{year}{2009}).

\bibitem[{\citenamefont{Bianchetti et~al.}(2010)\citenamefont{Bianchetti,
  Filipp, Baur, Fink, Lang, Steffen, Boissonneault, Blais, and
  Wallraff}}]{Bianchetti10}
\bibinfo{author}{\bibfnamefont{R.}~\bibnamefont{Bianchetti}},
  \bibinfo{author}{\bibfnamefont{S.}~\bibnamefont{Filipp}},
  \bibinfo{author}{\bibfnamefont{M.}~\bibnamefont{Baur}},
  \bibinfo{author}{\bibfnamefont{J.~M.} \bibnamefont{Fink}},
  \bibinfo{author}{\bibfnamefont{C.}~\bibnamefont{Lang}},
  \bibinfo{author}{\bibfnamefont{L.}~\bibnamefont{Steffen}},
  \bibinfo{author}{\bibfnamefont{M.}~\bibnamefont{Boissonneault}},
  \bibinfo{author}{\bibfnamefont{A.}~\bibnamefont{Blais}}, \bibnamefont{and}
  \bibinfo{author}{\bibfnamefont{A.}~\bibnamefont{Wallraff}},
  \bibinfo{journal}{Phys. Rev. Lett.} \textbf{\bibinfo{volume}{105}},
  \bibinfo{pages}{223601} (\bibinfo{year}{2010}).

\bibitem[{\citenamefont{Tian and Lloyd}(2000)}]{Tian2000}
\bibinfo{author}{\bibfnamefont{L.}~\bibnamefont{Tian}} \bibnamefont{and}
  \bibinfo{author}{\bibfnamefont{S.}~\bibnamefont{Lloyd}},
  \bibinfo{journal}{Phys. Rev. A} \textbf{\bibinfo{volume}{62}},
  \bibinfo{pages}{050301} (\bibinfo{year}{2000}).

\bibitem[{\citenamefont{Steffen et~al.}(2003)\citenamefont{Steffen, Martinis,
  and Chuang}}]{Steffen2003}
\bibinfo{author}{\bibfnamefont{M.}~\bibnamefont{Steffen}},
  \bibinfo{author}{\bibfnamefont{J.~M.} \bibnamefont{Martinis}},
  \bibnamefont{and} \bibinfo{author}{\bibfnamefont{I.~L.}
  \bibnamefont{Chuang}}, \bibinfo{journal}{Phys. Rev. B}
  \textbf{\bibinfo{volume}{89}}, \bibinfo{pages}{224518}
  (\bibinfo{year}{2003}).

\bibitem[{\citenamefont{Amin}(2006)}]{Amin2006}
\bibinfo{author}{\bibfnamefont{M.~H.~S.} \bibnamefont{Amin}},
  \bibinfo{journal}{Low Temp. Phys.} \textbf{\bibinfo{volume}{32}},
  \bibinfo{pages}{198} (\bibinfo{year}{2006}).

\bibitem[{\citenamefont{Strauch et~al.}(2007)\citenamefont{Strauch, Dutta,
  Paik, Palomaki, Mitra, Cooper, Lewis, Anderson, Dragt, Lobb
  et~al.}}]{Strauch2007}
\bibinfo{author}{\bibfnamefont{F.~W.} \bibnamefont{Strauch}},
  \bibinfo{author}{\bibfnamefont{S.~K.} \bibnamefont{Dutta}},
  \bibinfo{author}{\bibfnamefont{H.}~\bibnamefont{Paik}},
  \bibinfo{author}{\bibfnamefont{T.~A.} \bibnamefont{Palomaki}},
  \bibinfo{author}{\bibfnamefont{K.}~\bibnamefont{Mitra}},
  \bibinfo{author}{\bibfnamefont{B.~K.} \bibnamefont{Cooper}},
  \bibinfo{author}{\bibfnamefont{R.~M.} \bibnamefont{Lewis}},
  \bibinfo{author}{\bibfnamefont{J.~R.} \bibnamefont{Anderson}},
  \bibinfo{author}{\bibfnamefont{A.~J.} \bibnamefont{Dragt}},
  \bibinfo{author}{\bibfnamefont{C.~J.} \bibnamefont{Lobb}},
  \bibnamefont{et~al.}, \bibinfo{journal}{IEEE Trans. Appl. Supercond.}
  \textbf{\bibinfo{volume}{17}}, \bibinfo{pages}{105} (\bibinfo{year}{2007}).

\bibitem[{\citenamefont{Forney et~al.}(2010)\citenamefont{Forney, Jackson, and
  Strauch}}]{Forney10}
\bibinfo{author}{\bibfnamefont{A.~M.} \bibnamefont{Forney}},
  \bibinfo{author}{\bibfnamefont{S.~R.} \bibnamefont{Jackson}},
  \bibnamefont{and} \bibinfo{author}{\bibfnamefont{F.~W.}
  \bibnamefont{Strauch}}, \bibinfo{journal}{Phys. Rev. A}
  \textbf{\bibinfo{volume}{81}}, \bibinfo{pages}{012306}
  (\bibinfo{year}{2010}).

\bibitem[{\citenamefont{Motzoi et~al.}(2009)\citenamefont{Motzoi, Gambetta,
  Rebentrost, and Wilhelm}}]{Motzoi09}
\bibinfo{author}{\bibfnamefont{F.}~\bibnamefont{Motzoi}},
  \bibinfo{author}{\bibfnamefont{J.~M.} \bibnamefont{Gambetta}},
  \bibinfo{author}{\bibfnamefont{P.}~\bibnamefont{Rebentrost}},
  \bibnamefont{and} \bibinfo{author}{\bibfnamefont{F.~K.}
  \bibnamefont{Wilhelm}}, \bibinfo{journal}{Phys. Rev. Lett.}
  \textbf{\bibinfo{volume}{103}}, \bibinfo{pages}{110501}
  (\bibinfo{year}{2009}).

\bibitem[{\citenamefont{Lucero et~al.}(2010)\citenamefont{Lucero, Kelly,
  Bialczak, Lenander, Mariantoni, Neeley, O'Connell, Sank, Wang, Weides
  et~al.}}]{Lucero10}
\bibinfo{author}{\bibfnamefont{E.}~\bibnamefont{Lucero}},
  \bibinfo{author}{\bibfnamefont{J.}~\bibnamefont{Kelly}},
  \bibinfo{author}{\bibfnamefont{R.~C.} \bibnamefont{Bialczak}},
  \bibinfo{author}{\bibfnamefont{M.}~\bibnamefont{Lenander}},
  \bibinfo{author}{\bibfnamefont{M.}~\bibnamefont{Mariantoni}},
  \bibinfo{author}{\bibfnamefont{M.}~\bibnamefont{Neeley}},
  \bibinfo{author}{\bibfnamefont{A.~D.} \bibnamefont{O'Connell}},
  \bibinfo{author}{\bibfnamefont{D.}~\bibnamefont{Sank}},
  \bibinfo{author}{\bibfnamefont{H.}~\bibnamefont{Wang}},
  \bibinfo{author}{\bibfnamefont{M.}~\bibnamefont{Weides}},
  \bibnamefont{et~al.}, \bibinfo{journal}{Phys. Rev. A}
  \textbf{\bibinfo{volume}{82}}, \bibinfo{pages}{042339}
  (\bibinfo{year}{2010}).

\bibitem[{\citenamefont{Chow et~al.}(2010)\citenamefont{Chow, DiCarlo,
  Gambetta, Motzoi, Frunzio, Girvin, and Schoelkopf}}]{Chow10}
\bibinfo{author}{\bibfnamefont{J.~M.} \bibnamefont{Chow}},
  \bibinfo{author}{\bibfnamefont{L.}~\bibnamefont{DiCarlo}},
  \bibinfo{author}{\bibfnamefont{J.~M.} \bibnamefont{Gambetta}},
  \bibinfo{author}{\bibfnamefont{F.}~\bibnamefont{Motzoi}},
  \bibinfo{author}{\bibfnamefont{L.}~\bibnamefont{Frunzio}},
  \bibinfo{author}{\bibfnamefont{S.~M.} \bibnamefont{Girvin}},
  \bibnamefont{and} \bibinfo{author}{\bibfnamefont{R.~J.}
  \bibnamefont{Schoelkopf}}, \bibinfo{journal}{Phys. Rev. A}
  \textbf{\bibinfo{volume}{82}}, \bibinfo{pages}{040305}
  (\bibinfo{year}{2010}).

\bibitem[{\citenamefont{Hofheinz et~al.}(2008)\citenamefont{Hofheinz, Weig,
  Ansmann, Bialczak, Lucero, Neeley, O'Connell, Wang, Martinis, and
  Cleland}}]{Hofheinz2008}
\bibinfo{author}{\bibfnamefont{M.}~\bibnamefont{Hofheinz}},
  \bibinfo{author}{\bibfnamefont{E.~M.} \bibnamefont{Weig}},
  \bibinfo{author}{\bibfnamefont{M.}~\bibnamefont{Ansmann}},
  \bibinfo{author}{\bibfnamefont{R.~C.} \bibnamefont{Bialczak}},
  \bibinfo{author}{\bibfnamefont{E.}~\bibnamefont{Lucero}},
  \bibinfo{author}{\bibfnamefont{M.}~\bibnamefont{Neeley}},
  \bibinfo{author}{\bibfnamefont{A.~D.} \bibnamefont{O'Connell}},
  \bibinfo{author}{\bibfnamefont{H.}~\bibnamefont{Wang}},
  \bibinfo{author}{\bibfnamefont{J.~M.} \bibnamefont{Martinis}},
  \bibnamefont{and} \bibinfo{author}{\bibfnamefont{A.~N.}
  \bibnamefont{Cleland}}, \bibinfo{journal}{Nature}
  \textbf{\bibinfo{volume}{454}}, \bibinfo{pages}{310} (\bibinfo{year}{2008}).

\bibitem[{\citenamefont{Wang et~al.}(2008)\citenamefont{Wang, Hofheinz, Weig,
  Ansmann, Bialczak, Lucero, Neeley, O'Connell, Sank, Wenner
  et~al.}}]{Wang2008}
\bibinfo{author}{\bibfnamefont{H.}~\bibnamefont{Wang}},
  \bibinfo{author}{\bibfnamefont{M.}~\bibnamefont{Hofheinz}},
  \bibinfo{author}{\bibfnamefont{E.~M.} \bibnamefont{Weig}},
  \bibinfo{author}{\bibfnamefont{M.}~\bibnamefont{Ansmann}},
  \bibinfo{author}{\bibfnamefont{R.~C.} \bibnamefont{Bialczak}},
  \bibinfo{author}{\bibfnamefont{E.}~\bibnamefont{Lucero}},
  \bibinfo{author}{\bibfnamefont{M.}~\bibnamefont{Neeley}},
  \bibinfo{author}{\bibfnamefont{A.~D.} \bibnamefont{O'Connell}},
  \bibinfo{author}{\bibfnamefont{D.}~\bibnamefont{Sank}},
  \bibinfo{author}{\bibfnamefont{J.}~\bibnamefont{Wenner}},
  \bibnamefont{et~al.}, \bibinfo{journal}{Phys. Rev. Lett.}
  \textbf{\bibinfo{volume}{101}}, \bibinfo{pages}{240401}
  (\bibinfo{year}{2008}).

\bibitem[{\citenamefont{Hofheinz et~al.}(2009)\citenamefont{Hofheinz, Wang,
  Ansmann, Bialczak, Lucero, Neeley, O'Connell, Sank, Wenner, Martinis
  et~al.}}]{Hofheinz2009}
\bibinfo{author}{\bibfnamefont{M.}~\bibnamefont{Hofheinz}},
  \bibinfo{author}{\bibfnamefont{H.}~\bibnamefont{Wang}},
  \bibinfo{author}{\bibfnamefont{M.}~\bibnamefont{Ansmann}},
  \bibinfo{author}{\bibfnamefont{R.~C.} \bibnamefont{Bialczak}},
  \bibinfo{author}{\bibfnamefont{E.}~\bibnamefont{Lucero}},
  \bibinfo{author}{\bibfnamefont{M.}~\bibnamefont{Neeley}},
  \bibinfo{author}{\bibfnamefont{A.~D.} \bibnamefont{O'Connell}},
  \bibinfo{author}{\bibfnamefont{D.}~\bibnamefont{Sank}},
  \bibinfo{author}{\bibfnamefont{J.}~\bibnamefont{Wenner}},
  \bibinfo{author}{\bibfnamefont{J.~M.} \bibnamefont{Martinis}},
  \bibnamefont{et~al.}, \bibinfo{journal}{Nature}
  \textbf{\bibinfo{volume}{459}}, \bibinfo{pages}{456} (\bibinfo{year}{2009}).

\bibitem[{\citenamefont{Law and Eberly}(1996)}]{Law96}
\bibinfo{author}{\bibfnamefont{C.~K.} \bibnamefont{Law}} \bibnamefont{and}
  \bibinfo{author}{\bibfnamefont{J.~H.} \bibnamefont{Eberly}},
  \bibinfo{journal}{Phys. Rev. Lett.} \textbf{\bibinfo{volume}{76}},
  \bibinfo{pages}{1055} (\bibinfo{year}{1996}).

\bibitem[{\citenamefont{Strauch et~al.}(2010)\citenamefont{Strauch, Jacobs, and
  Simmonds}}]{Strauch10}
\bibinfo{author}{\bibfnamefont{F.~W.} \bibnamefont{Strauch}},
  \bibinfo{author}{\bibfnamefont{K.}~\bibnamefont{Jacobs}}, \bibnamefont{and}
  \bibinfo{author}{\bibfnamefont{R.~W.} \bibnamefont{Simmonds}},
  \bibinfo{journal}{Phys. Rev. Lett.} \textbf{\bibinfo{volume}{105}},
  \bibinfo{pages}{050501} (\bibinfo{year}{2010}).

\bibitem[{\citenamefont{Wang et~al.}(2011)\citenamefont{Wang, Mariantoni,
  Bialczak, Lenander, Lucero, Neeley, O'Connell, Sank, Weides, Wenner
  et~al.}}]{wang11}
\bibinfo{author}{\bibfnamefont{H.}~\bibnamefont{Wang}},
  \bibinfo{author}{\bibfnamefont{M.}~\bibnamefont{Mariantoni}},
  \bibinfo{author}{\bibfnamefont{R.~C.} \bibnamefont{Bialczak}},
  \bibinfo{author}{\bibfnamefont{M.}~\bibnamefont{Lenander}},
  \bibinfo{author}{\bibfnamefont{E.}~\bibnamefont{Lucero}},
  \bibinfo{author}{\bibfnamefont{M.}~\bibnamefont{Neeley}},
  \bibinfo{author}{\bibfnamefont{A.~D.} \bibnamefont{O'Connell}},
  \bibinfo{author}{\bibfnamefont{D.}~\bibnamefont{Sank}},
  \bibinfo{author}{\bibfnamefont{M.}~\bibnamefont{Weides}},
  \bibinfo{author}{\bibfnamefont{J.}~\bibnamefont{Wenner}},
  \bibnamefont{et~al.}, \bibinfo{journal}{Phys. Rev. Lett.}
  \textbf{\bibinfo{volume}{106}}, \bibinfo{pages}{060401}
  (\bibinfo{year}{2011}).

\bibitem[{\citenamefont{Dowling}(2008)}]{Dowling08}
\bibinfo{author}{\bibfnamefont{J.~P.} \bibnamefont{Dowling}},
  \bibinfo{journal}{Contemp. Phys.} \textbf{\bibinfo{volume}{49}},
  \bibinfo{pages}{125} (\bibinfo{year}{2008}).

\bibitem[{\citenamefont{Merkel and Wilhelm}(2010)}]{Merkel10}
\bibinfo{author}{\bibfnamefont{S.~T.} \bibnamefont{Merkel}} \bibnamefont{and}
  \bibinfo{author}{\bibfnamefont{F.~K.} \bibnamefont{Wilhelm}},
  \bibinfo{journal}{New Journal of Physics} \textbf{\bibinfo{volume}{12}},
  \bibinfo{pages}{093036} (\bibinfo{year}{2010}).

\bibitem[{\citenamefont{Nielsen and Chuang}(2000)}]{mikeandike}
\bibinfo{author}{\bibfnamefont{M.~A.} \bibnamefont{Nielsen}} \bibnamefont{and}
  \bibinfo{author}{\bibfnamefont{I.~L.} \bibnamefont{Chuang}},
  \emph{\bibinfo{title}{Quantum Computation and Quantum Information}}
  (\bibinfo{publisher}{Cambridge University Press}, \bibinfo{year}{2000}).

\bibitem[{\citenamefont{Jacobs}(2007)}]{Jacobs07x}
\bibinfo{author}{\bibfnamefont{K.}~\bibnamefont{Jacobs}},
  \bibinfo{journal}{Phys. Rev. Lett.} \textbf{\bibinfo{volume}{99}},
  \bibinfo{pages}{117203} (\bibinfo{year}{2007}).

\bibitem[{\citenamefont{Lloyd et~al.}(2004)\citenamefont{Lloyd, Landahl, and
  Slotine}}]{Lloyd04}
\bibinfo{author}{\bibfnamefont{S.}~\bibnamefont{Lloyd}},
  \bibinfo{author}{\bibfnamefont{A.~J.} \bibnamefont{Landahl}},
  \bibnamefont{and} \bibinfo{author}{\bibfnamefont{J.-J.~E.}
  \bibnamefont{Slotine}}, \bibinfo{journal}{Phys. Rev. A}
  \textbf{\bibinfo{volume}{69}}, \bibinfo{pages}{012305}
  (\bibinfo{year}{2004}).

\bibitem[{\citenamefont{Jacobs and Landahl}(2009)}]{Jacobs09c}
\bibinfo{author}{\bibfnamefont{K.}~\bibnamefont{Jacobs}} \bibnamefont{and}
  \bibinfo{author}{\bibfnamefont{A.~J.} \bibnamefont{Landahl}},
  \bibinfo{journal}{Phys. Rev. Lett.} \textbf{\bibinfo{volume}{103}},
  \bibinfo{pages}{067201} (\bibinfo{year}{2009}).

\bibitem[{\citenamefont{Santos}(2005)}]{Santos05}
\bibinfo{author}{\bibfnamefont{M.~F.} \bibnamefont{Santos}},
  \bibinfo{journal}{Phys. Rev. Lett.} \textbf{\bibinfo{volume}{95}},
  \bibinfo{pages}{010504} (\bibinfo{year}{2005}).

\bibitem[{\citenamefont{Schuster et~al.}(2007)\citenamefont{Schuster, Houck,
  Schreier, Wallraff, Gambetta, Blais, Frunzio, Majer, Johnson, Devoret
  et~al.}}]{Schuster07}
\bibinfo{author}{\bibfnamefont{D.~I.} \bibnamefont{Schuster}},
  \bibinfo{author}{\bibfnamefont{A.~A.} \bibnamefont{Houck}},
  \bibinfo{author}{\bibfnamefont{J.~A.} \bibnamefont{Schreier}},
  \bibinfo{author}{\bibfnamefont{A.}~\bibnamefont{Wallraff}},
  \bibinfo{author}{\bibfnamefont{J.~M.} \bibnamefont{Gambetta}},
  \bibinfo{author}{\bibfnamefont{A.}~\bibnamefont{Blais}},
  \bibinfo{author}{\bibfnamefont{L.}~\bibnamefont{Frunzio}},
  \bibinfo{author}{\bibfnamefont{J.}~\bibnamefont{Majer}},
  \bibinfo{author}{\bibfnamefont{B.}~\bibnamefont{Johnson}},
  \bibinfo{author}{\bibfnamefont{M.~H.} \bibnamefont{Devoret}},
  \bibnamefont{et~al.}, \bibinfo{journal}{Nature}
  \textbf{\bibinfo{volume}{445}}, \bibinfo{pages}{515} (\bibinfo{year}{2007}).

\bibitem[{\citenamefont{Johnson et~al.}(2010)\citenamefont{Johnson, Reed,
  Houck, Schuster, Bishop, Ginossar, Gambetta, DiCarlo, Frunzio, and
  {\textit{et al.}}}}]{Johnson10}
\bibinfo{author}{\bibfnamefont{B.~R.} \bibnamefont{Johnson}},
  \bibinfo{author}{\bibfnamefont{M.~D.} \bibnamefont{Reed}},
  \bibinfo{author}{\bibfnamefont{A.~A.} \bibnamefont{Houck}},
  \bibinfo{author}{\bibfnamefont{D.~I.} \bibnamefont{Schuster}},
  \bibinfo{author}{\bibfnamefont{L.~S.} \bibnamefont{Bishop}},
  \bibinfo{author}{\bibfnamefont{E.}~\bibnamefont{Ginossar}},
  \bibinfo{author}{\bibfnamefont{J.~M.} \bibnamefont{Gambetta}},
  \bibinfo{author}{\bibfnamefont{L.}~\bibnamefont{DiCarlo}},
  \bibinfo{author}{\bibfnamefont{L.}~\bibnamefont{Frunzio}}, \bibnamefont{and}
  \bibinfo{author}{\bibfnamefont{S.~M.~G.} \bibnamefont{{\textit{et al.}}}},
  \bibinfo{journal}{Nature Physics} \textbf{\bibinfo{volume}{6}},
  \bibinfo{pages}{663} (\bibinfo{year}{2010}).

\bibitem[{\citenamefont{Gottesman}(1999)}]{Gottesman99}
\bibinfo{author}{\bibfnamefont{D.}~\bibnamefont{Gottesman}},
  \bibinfo{journal}{Chaos, Solitons \& Fractals} \textbf{\bibinfo{volume}{10}},
  \bibinfo{pages}{1749} (\bibinfo{year}{1999}).

\bibitem[{\citenamefont{Gottesman et~al.}(2001)\citenamefont{Gottesman, Kitaev,
  and Preskill}}]{Gottesman01}
\bibinfo{author}{\bibfnamefont{D.}~\bibnamefont{Gottesman}},
  \bibinfo{author}{\bibfnamefont{A.}~\bibnamefont{Kitaev}}, \bibnamefont{and}
  \bibinfo{author}{\bibfnamefont{J.}~\bibnamefont{Preskill}},
  \bibinfo{journal}{Phys. Rev. A} \textbf{\bibinfo{volume}{64}},
  \bibinfo{pages}{012310} (\bibinfo{year}{2001}).

\bibitem[{\citenamefont{Muthukrishnan and Stroud}(2000)}]{Stroud2000}
\bibinfo{author}{\bibfnamefont{A.}~\bibnamefont{Muthukrishnan}}
  \bibnamefont{and} \bibinfo{author}{\bibfnamefont{C.~R.}
  \bibnamefont{Stroud}}, \bibinfo{journal}{Phys. Rev. A}
  \textbf{\bibinfo{volume}{62}}, \bibinfo{pages}{052309}
  (\bibinfo{year}{2000}).

\bibitem[{\citenamefont{Bartlett et~al.}(2002)\citenamefont{Bartlett, {de
  Guise}, and Sanders}}]{Bartlett02}
\bibinfo{author}{\bibfnamefont{S.~D.} \bibnamefont{Bartlett}},
  \bibinfo{author}{\bibfnamefont{H.}~\bibnamefont{{de Guise}}},
  \bibnamefont{and} \bibinfo{author}{\bibfnamefont{B.~C.}
  \bibnamefont{Sanders}}, \bibinfo{journal}{Phys. Rev. A}
  \textbf{\bibinfo{volume}{65}}, \bibinfo{pages}{052316}
  (\bibinfo{year}{2002}).

\bibitem[{\citenamefont{Brennen et~al.}(2005)\citenamefont{Brennen, O'Leary,
  and Bullock}}]{Brennen05}
\bibinfo{author}{\bibfnamefont{G.~K.} \bibnamefont{Brennen}},
  \bibinfo{author}{\bibfnamefont{D.~P.} \bibnamefont{O'Leary}},
  \bibnamefont{and} \bibinfo{author}{\bibfnamefont{S.~S.}
  \bibnamefont{Bullock}}, \bibinfo{journal}{Phys. Rev. A}
  \textbf{\bibinfo{volume}{71}}, \bibinfo{pages}{052318}
  (\bibinfo{year}{2005}).

\bibitem[{\citenamefont{Bullock et~al.}(2005)\citenamefont{Bullock, O'Leary,
  and Brennen}}]{Bullock05}
\bibinfo{author}{\bibfnamefont{S.~S.} \bibnamefont{Bullock}},
  \bibinfo{author}{\bibfnamefont{D.~P.} \bibnamefont{O'Leary}},
  \bibnamefont{and} \bibinfo{author}{\bibfnamefont{G.~K.}
  \bibnamefont{Brennen}}, \bibinfo{journal}{Phys. Rev. Lett.}
  \textbf{\bibinfo{volume}{94}}, \bibinfo{pages}{230502}
  (\bibinfo{year}{2005}).

\bibitem[{\citenamefont{O'Leary et~al.}(2006)\citenamefont{O'Leary, Brennen,
  and Bullock}}]{OLeary06}
\bibinfo{author}{\bibfnamefont{D.~P.} \bibnamefont{O'Leary}},
  \bibinfo{author}{\bibfnamefont{G.~K.} \bibnamefont{Brennen}},
  \bibnamefont{and} \bibinfo{author}{\bibfnamefont{S.~S.}
  \bibnamefont{Bullock}}, \bibinfo{journal}{Phys. Rev. A.}
  \textbf{\bibinfo{volume}{74}}, \bibinfo{pages}{032334}
  (\bibinfo{year}{2006}).

\bibitem[{\citenamefont{Allman et~al.}(2010)\citenamefont{Allman, Altomare,
  Whittaker, Cicak, Li, Sirois, Teufel, and Simmonds}}]{Allman2010}
\bibinfo{author}{\bibfnamefont{M.}~\bibnamefont{Allman}},
  \bibinfo{author}{\bibfnamefont{F.}~\bibnamefont{Altomare}},
  \bibinfo{author}{\bibfnamefont{J.~D.} \bibnamefont{Whittaker}},
  \bibinfo{author}{\bibfnamefont{K.}~\bibnamefont{Cicak}},
  \bibinfo{author}{\bibfnamefont{D.}~\bibnamefont{Li}},
  \bibinfo{author}{\bibfnamefont{A.}~\bibnamefont{Sirois}},
  \bibinfo{author}{\bibfnamefont{J.~D.} \bibnamefont{Teufel}},
  \bibnamefont{and} \bibinfo{author}{\bibfnamefont{R.~W.}
  \bibnamefont{Simmonds}}, \bibinfo{journal}{Phys. Rev. Lett.}
  \textbf{\bibinfo{volume}{104}}, \bibinfo{pages}{177004}
  (\bibinfo{year}{2010}).

\bibitem[{\citenamefont{Bialczak et~al.}(2010)\citenamefont{Bialczak, Ansmann,
  Hofheinz, Lenander, Lucero, Neeley, O'Connell, Sank, Wang, Weides
  et~al.}}]{Bialczak10}
\bibinfo{author}{\bibfnamefont{R.~C.} \bibnamefont{Bialczak}},
  \bibinfo{author}{\bibfnamefont{M.}~\bibnamefont{Ansmann}},
  \bibinfo{author}{\bibfnamefont{M.}~\bibnamefont{Hofheinz}},
  \bibinfo{author}{\bibfnamefont{M.}~\bibnamefont{Lenander}},
  \bibinfo{author}{\bibfnamefont{E.}~\bibnamefont{Lucero}},
  \bibinfo{author}{\bibfnamefont{M.}~\bibnamefont{Neeley}},
  \bibinfo{author}{\bibfnamefont{A.}~\bibnamefont{O'Connell}},
  \bibinfo{author}{\bibfnamefont{D.}~\bibnamefont{Sank}},
  \bibinfo{author}{\bibfnamefont{H.}~\bibnamefont{Wang}},
  \bibinfo{author}{\bibfnamefont{M.}~\bibnamefont{Weides}},
  \bibnamefont{et~al.}, \bibinfo{journal}{arXiv:1007.2219}
  (\bibinfo{year}{2010}).

\bibitem[{\citenamefont{Koch et~al.}(2007)\citenamefont{Koch, Yu, Gambetta,
  Houck, Schuster, Majer, Blais, Devoret, Girvin, and Schoelkopf}}]{Koch07}
\bibinfo{author}{\bibfnamefont{J.}~\bibnamefont{Koch}},
  \bibinfo{author}{\bibfnamefont{T.~M.} \bibnamefont{Yu}},
  \bibinfo{author}{\bibfnamefont{J.}~\bibnamefont{Gambetta}},
  \bibinfo{author}{\bibfnamefont{A.~A.} \bibnamefont{Houck}},
  \bibinfo{author}{\bibfnamefont{D.~I.} \bibnamefont{Schuster}},
  \bibinfo{author}{\bibfnamefont{J.}~\bibnamefont{Majer}},
  \bibinfo{author}{\bibfnamefont{A.}~\bibnamefont{Blais}},
  \bibinfo{author}{\bibfnamefont{M.~H.} \bibnamefont{Devoret}},
  \bibinfo{author}{\bibfnamefont{S.~M.} \bibnamefont{Girvin}},
  \bibnamefont{and} \bibinfo{author}{\bibfnamefont{R.~J.}
  \bibnamefont{Schoelkopf}}, \bibinfo{journal}{Phys. Rev. A}
  \textbf{\bibinfo{volume}{76}}, \bibinfo{pages}{042319}
  (\bibinfo{year}{2007}).

\bibitem[{\citenamefont{Sillanp{\"a\"a}
  et~al.}(2007)\citenamefont{Sillanp{\"a\"a}, Park, and
  Simmonds}}]{Sillanpaa2007}
\bibinfo{author}{\bibfnamefont{M.}~\bibnamefont{Sillanp{\"a\"a}}},
  \bibinfo{author}{\bibfnamefont{J.~I.} \bibnamefont{Park}}, \bibnamefont{and}
  \bibinfo{author}{\bibfnamefont{R.~W.} \bibnamefont{Simmonds}},
  \bibinfo{journal}{Nature} \textbf{\bibinfo{volume}{449}},
  \bibinfo{pages}{438} (\bibinfo{year}{2007}).

\bibitem[{\citenamefont{Johnson et~al.}(2003)\citenamefont{Johnson, Strauch,
  Dragt, Ramos, Lobb, Anderson, and Wellstood}}]{Johnson2003}
\bibinfo{author}{\bibfnamefont{P.~R.} \bibnamefont{Johnson}},
  \bibinfo{author}{\bibfnamefont{F.~W.} \bibnamefont{Strauch}},
  \bibinfo{author}{\bibfnamefont{A.~J.} \bibnamefont{Dragt}},
  \bibinfo{author}{\bibfnamefont{R.~C.} \bibnamefont{Ramos}},
  \bibinfo{author}{\bibfnamefont{C.~J.} \bibnamefont{Lobb}},
  \bibinfo{author}{\bibfnamefont{J.~R.} \bibnamefont{Anderson}},
  \bibnamefont{and} \bibinfo{author}{\bibfnamefont{F.~C.}
  \bibnamefont{Wellstood}}, \bibinfo{journal}{Phys.\ Rev.\ B}
  \textbf{\bibinfo{volume}{67}}, \bibinfo{pages}{020509}
  (\bibinfo{year}{2003}).

\bibitem[{\citenamefont{Mariantoni et~al.}(2011)\citenamefont{Mariantoni, Wang,
  Yamamoto, Neeley, Bialczak, Chen, Lenander, Lucero, O'Connell, Sank
  et~al.}}]{Mariantoni11}
\bibinfo{author}{\bibfnamefont{M.}~\bibnamefont{Mariantoni}},
  \bibinfo{author}{\bibfnamefont{H.}~\bibnamefont{Wang}},
  \bibinfo{author}{\bibfnamefont{T.}~\bibnamefont{Yamamoto}},
  \bibinfo{author}{\bibfnamefont{M.}~\bibnamefont{Neeley}},
  \bibinfo{author}{\bibfnamefont{R.}~\bibnamefont{Bialczak}},
  \bibinfo{author}{\bibfnamefont{Y.}~\bibnamefont{Chen}},
  \bibinfo{author}{\bibfnamefont{M.}~\bibnamefont{Lenander}},
  \bibinfo{author}{\bibfnamefont{E.}~\bibnamefont{Lucero}},
  \bibinfo{author}{\bibfnamefont{A.~D.} \bibnamefont{O'Connell}},
  \bibinfo{author}{\bibfnamefont{D.}~\bibnamefont{Sank}}, \bibnamefont{et~al.},
  \bibinfo{journal}{Science} \textbf{\bibinfo{volume}{334}},
  \bibinfo{pages}{61} (\bibinfo{year}{2011}).

\bibitem[{\citenamefont{Paik et~al.}(2011)\citenamefont{Paik, Schuster, Bishop,
  Kirchmair, Catelani, Sears, Johnson, Reagor, Frunzio, Glazman
  et~al.}}]{Paik2011}
\bibinfo{author}{\bibfnamefont{H.}~\bibnamefont{Paik}},
  \bibinfo{author}{\bibfnamefont{D.~I.} \bibnamefont{Schuster}},
  \bibinfo{author}{\bibfnamefont{L.~S.} \bibnamefont{Bishop}},
  \bibinfo{author}{\bibfnamefont{G.}~\bibnamefont{Kirchmair}},
  \bibinfo{author}{\bibfnamefont{G.}~\bibnamefont{Catelani}},
  \bibinfo{author}{\bibfnamefont{A.~P.} \bibnamefont{Sears}},
  \bibinfo{author}{\bibfnamefont{B.~R.} \bibnamefont{Johnson}},
  \bibinfo{author}{\bibfnamefont{M.~J.} \bibnamefont{Reagor}},
  \bibinfo{author}{\bibfnamefont{L.}~\bibnamefont{Frunzio}},
  \bibinfo{author}{\bibfnamefont{L.}~\bibnamefont{Glazman}},
  \bibnamefont{et~al.}, \bibinfo{journal}{eprint: arXiv:1105.4642}
  (\bibinfo{year}{2011}).

\bibitem[{\citenamefont{Merkel et~al.}(2009)\citenamefont{Merkel, Jessen, and
  Deutsch}}]{Merkel09}
\bibinfo{author}{\bibfnamefont{S.~T.} \bibnamefont{Merkel}},
  \bibinfo{author}{\bibfnamefont{G.~B. P.~S.} \bibnamefont{Jessen}},
  \bibnamefont{and} \bibinfo{author}{\bibfnamefont{I.~H.}
  \bibnamefont{Deutsch}}, \bibinfo{journal}{Phys. Rev. A}
  \textbf{\bibinfo{volume}{80}}, \bibinfo{pages}{023424}
  (\bibinfo{year}{2009}).

\end{thebibliography}

\end{document}